# Transient gas-liquid flow phenomena in M-shaped jumper of subsea gas production systems during start-up operation

A. Yurishchev, R.B. Ravid, A. Ullmann, N. Brauner

*School of Mechanical Engineering, Tel-Aviv University, Tel-Aviv 6997801, Israel*

## Abstract

Removal of the accumulated liquid from jumpers of subsea gas production systems is essential to avoid possible hydrate creation and further damage to the pipeline. However, the displacement of high amounts of accumulated liquid during the production start-up leads to a gas pressure rise. Liquid plugs formed during the liquid displacement impact the structure's elbows. This, in addition to cyclic pressure/forces fluctuations, may lead to harmful flow-induced vibrations (FIV). These flow phenomena that may endanger the jumper structure were explored in air-water experiments performed in a lab-scale jumper. The critical (minimal) gas velocity needed to purge the accumulated liquid was determined and the pressure and forces variations during the liquid removal were measured. In addition, the effects of the gas velocity, initial liquid amount, and gas flow ramp-up on the air-water flow phenomena were documented. Results of 3D and 2D numerical simulations (using OpenFOAM) were verified against the experimental data. The effects of employing different RANS turbulence models on the predictions were tested and demonstrated. A simple mechanistic model was established to predict the pressure and force variation during liquid displacement. The model enables inspecting the variation with the operational conditions of each pressure component (i.e., hydrostatic, friction, and acceleration) and examining their significance and contribution to the pressure rise.

**Keywords:** natural gas, subsea jumper, accumulated liquid removal, critical gas velocity, pressure rise, forces on elbows

# 1 Introduction

Subsea jumpers are widely used in deep-water offshore gas production systems. A jumper is a tie-in system that is used to connect two subsea production structures (e.g., the wellhead to the Infield Gathering Manifold (IMF), the IMF to the export flow lines via their respective PLET, i.e., Pipeline End Termination). Although flexible jumpers are available, rigid jumpers are commonly used. The jumper parts can be assembled into a variety of spatial configurations, such as U-shape (vertical or horizontal), inverted U-shape, M-shape, S-shape, L-shape, Z-shape, or Arc, by using steel bends, tees, and elbows as necessary. The design of all parts of a rigid jumper system should consider aspects related to reliability and safe operation to minimize failures and maintenance problems. Thus, jumpers are built to accommodate high static and dynamic loads due to thermal expansion, high internal pressure, and hydrodynamic loads from the internal multiphase flow and external currents.

In offshore natural-gas production systems, two-phase flow is encountered in the subsea gas transportation lines due to the presence of liquids. A liquid phase may originate from the presence of condensates and some water from the reservoir. The latter becomes significant, mainly in the late stage of the production. The liquid phase also originates from the chemicals added to prevent hydrates formation, such as methanol or mono-ethylene glycol (MEG) solutions (e.g., Hemmingsen et al., 2011; Kim et al., 2017; Lim et al., 2022; Liu et al., 2021).

Natural gas transportation pipelines in general (and jumpers in particular) usually have low spot sections, where the liquids can accumulate, especially during operation shutdowns (e.g., Birvalski et al., 2014; Hamami Bissor (Abir) et al., 2021). Upon restart, the gas displaces the accumulated liquid, but the flow can temporarily be associated with flow patterns typical to high liquid holdups (e.g., slug flow). Liquid accumulation in the jumper's low sections may also be encountered at continuous operation when the gas production rate gets below a threshold flow rate required to ensure continuous removal of the liquid from the jumpers.

Critical superficial gas velocity needed for propelling the liquid upwards in an inclined pipe was reported in previous studies. For instance, in long slightly inclined pipes, the liquid will usually propagate as a thin film and the critical gas velocity will satisfy the avoidance of liquid backflow near the wall (e.g., Fan et al., 2018; Hamami Bissor et al., 2020; Soedarmo et al., 2019; Wang et al., 2016). In vertical or off-vertical pipes, the critical gas velocity for establishing stable upward gas-liquid flow at low liquid loads is related to the transition to annular flow and the avoidance of flow reversal of the liquid in the annulus (e.g., Liu et al., 2018; Vieira et al., 2021; Waltrich et al., 2015a, 2015b; Wang et al., 2022). Those mechanisms can be relevant for predicting the critical gas velocity in the riser section of jumpers.

The two-phase flow patterns associated with the liquid displacement by the gas in jumpers depend on the various system parameters, including the gas pressure and flow rate, the amount of liquid present in the system and the jumper shape and size. Consequently, different flow patterns may be encountered under steady and transient operation modes. Each of the flow patterns is associated with a range of pressure fluctuation frequencies, which interact with the jumper structure and may induce harmful Flow-Induced Vibration (FIV). Even low cyclic stresses pose structural integrity concerns since, over time, they may lead to fatigue failure. Moreover, a dominant fluctuating frequency that matches the structure's natural frequency can result in a resonance, where large vibrations of the structure occur in response to small amplitude force fluctuations.

Understanding the consequences of internal two-phase FIV is crucial for securing the reliability and integrity of gas transportation systems. However, vibrations due to the internal flow have received less attention compared to those induced by external flows. The effects of the external subsea currents on jumper structures were investigated both experimentally (e.g., Amini and Fernandes, 2015; Gross et al., 2018) and numerically (e.g., Holmes and Constantinides, 2010; Qu et al., 2022; Zheng et al., 2015). These studies revealed the natural frequencies of the structures along with VIV (Vortex Induced Vibrations) frequencies under various external flow conditions. Some studies on jumpers also focused on the internal two-phase flow dynamics considering continuous operation and its effects on the real scale jumper structures (FSI - Fluid Structure Interaction with either one-way or two-way coupling (e.g., Elyyan et al., 2014; Kim and Srinil, 2018; Nair et al., 2011).

In a more recent study, Li et al., 2022 combined experiments and numerical simulations (one-way FSI) on a multi-planar (Z-shaped) jumper of ID=48mm pipes at high-pressure conditions (20 bar). Numerical simulations qualitatively predicted the flow patterns for the tested cases, and sufficient agreement between the simulated and experimental structure vibration frequencies was reported. The structure vibration characteristics were found to be related to the two-phase flow patterns. Also, the vibration frequencies increase with the increase of the liquid superficial velocity. Zhu et al., 2023, 2022 extended the numerical investigations to M- and reversed U-shaped jumpers.

Several studies addressed the impact of liquid plugs/slugs on elbows and bends (e.g., Garcia et al., 2023; Riverin et al., 2006). Bamidele et al., 2022 investigated experimentally the structure displacement and forces applied on the bends for reversed U-shaped scaled-down (ID=1 inch) jumper. Effects of the two-phase flow properties on the vibration amplitudes were reported. In addition, a model for the prediction of the impact forces on a bend due to slug flow was introduced, considering a realistic slug flow case. The suggested Slug Flow Model (SFL) showed an improvement in the prediction of the maximal impact force compared to a similar Piston Flow Model (PFL) of Tay and Thorpe, 2014. The latter model predictions tend to somewhat overestimate the experimental results.

The literature review indicates that most previous studies focused on continuous slug flow conditions in jumpers, which are associated with high liquid load in the produced gas. However, the gas-liquid dynamics during transient conditions were not yet examined, in particular during start–up of the gas production following a shutdown, when initially a certain volume of liquid rests at lower sections. During start-up, the gas flow rate should reach a steady operational production rate, which should be sufficiently high to purge out large amounts of accumulated liquids. While in regular operation, the gas flow lines are typically associated with low liquid load, this may not be the case during the start-up period and during transients.

In this study, we report experimental data and numerical simulations associated with transient operational conditions that aimed at (1) identifying the minimal (critical) gas velocity needed to prevent backflow in the riser and to remove the residual liquid from the riser's low elbow; (2) detecting the internal pressure variations and peaks associated with the liquid removal; (3) reporting the loads (forces) applied on the jumper's elbows during the rapid liquid displacement. A mechanistic model is proposed to predict the pressure variation and peak during the liquid purge and to estimate the forces acting on the jumper's elbows.

# 2 Experimental setup

The flow loop and the test rig (see **Figure 2.1**) were designed to mimic a small-scale rigid M-shaped typical subsea jumper configuration. The test rig is made of transparent Perspex pipes of $D$=50 mm ID, including the elbows with a radius of curvature of 3$D$ (=150mm). The lengths of the test sections are 70$D$ for the horizontal section and 40$D$/70$D$ for the down-comer/upcomer (riser) sections. The jumper lab model is mounted on a supportive construction made of aluminum bars and rigid connectors. The transparent pipes enable observation of the flow phenomena. A high-speed video camera (Photron FASTCAM Mini AX50) and an illumination system were used for the documentation of the two-phase flow characteristics along the test sections.

A centrifugal pump (GUANGDONG YUEHUA GZA32-160/3) with controllable r.p.m (by Delta VFD-EL) is used to introduce the precise amount of the water (measured by electromagnetic flow meter ISOMAG MS 100) into the system. The air flow rate is controlled by a mass flow meter and controller (Alicat-MCR Mass Flow Controller series). The controller allows a maximum volumetric flow of 3000SLPM, accuracy of ±0.2% of the full scale. The full range of the air flow rates (up to 3000 SLPM, corresponding to air superficial velocity of 0-25 m/s) was tested, and the experimental runs were repeated for a different prefilled amount of water. The gas flow rate was gradually increased, following a linear set point ramp-up of ~2, 4 and 6 seconds until stabilization of the flow rate at the final desired value was achieved.

In each experimental run, the transient pressure drop on the jumper (between the inlet and the point located downstream to the riser's upper elbow) was monitored by a pressure transducer (SETRA Model 231RS) in order to measure the maximum peak and the following pressure fluctuations. In addition, a quartz ICP© impact force sensor (model 208C02) was used to document the horizontal force exerted on the riser's upper elbow when the water passed through. Measurement errors of mass-flow controller, pressure transducer and impact force meter are provided in **Table 2.1**. The sampling rate of the pressure transducer and the impact force sensor is 100 Hz.

To enable measurement of the force exerted by the flow on the elbow, the latter was connected to the jumper with flexible joints. To verify that these joints do not affect the measured force, we applied an external force to the pipe sections connected to the elbow upstream and downstream of the flexible connections. No force was detected by the mounted force transducer. In fact, the elbow is connected to the supporting structure through the force transducer. Hence, the elbow displacement during the experiment is negligible and does not affect the flow phenomena.

Experiments with different prefilled amounts of water and varying air flow rates were performed to characterize the two-phase flow patterns and to determine the minimum air velocity needed to remove the entire liquid from the pipeline. The air-water two-phase flow dynamics were captured using a high-speed camera (250-500 FPS) for a range of gas velocities. The flow visualization test section is located at ~20D downstream of the riser's lower elbow, where the effect of the bend on the flow dynamics appears negligible. The flow in the transparent riser's lower elbow was also documented and examined. In both cases, a white opaque screen is placed behind the test section, and the screen is illuminated by a set of LEDs, in order to allow capturing a clear and sharp picture.

The wetting properties (contact angles with wetted Perspex surface) were determined experimentally. To this aim, an OCA 15EC (Optical Contact Angle) apparatus combined with TBU 100 (Tiltable Base Unit) was utilized (SDM-Sessile Drop Method). The measured static, advancing and receding contact angle values on a used (pre-wetted) surface are in the range of 65°-70°, 70°-75° and 30°-40°, respectively.

**Table 2.1*:*** Measurement errors of mass-flow controller, pressure transducer and impact force meter.

| ***Device*** | ***Full Scale*** | ***Measurement Error*** |
|---|---|---|
| Mass Flow Controller: Alicat MCR-Series | 3000 SLPM ≈ 25.5 m/s | 0.2% FS |
| Pressure Transducer: SETRA Model 231RS | 10 PSI ≈ 69 kPa | 1% FS |
| Impact Force Sensor Meter: ICP© Model 208C02 | 100 lb ≈ 0.45 kN | 1% FS |

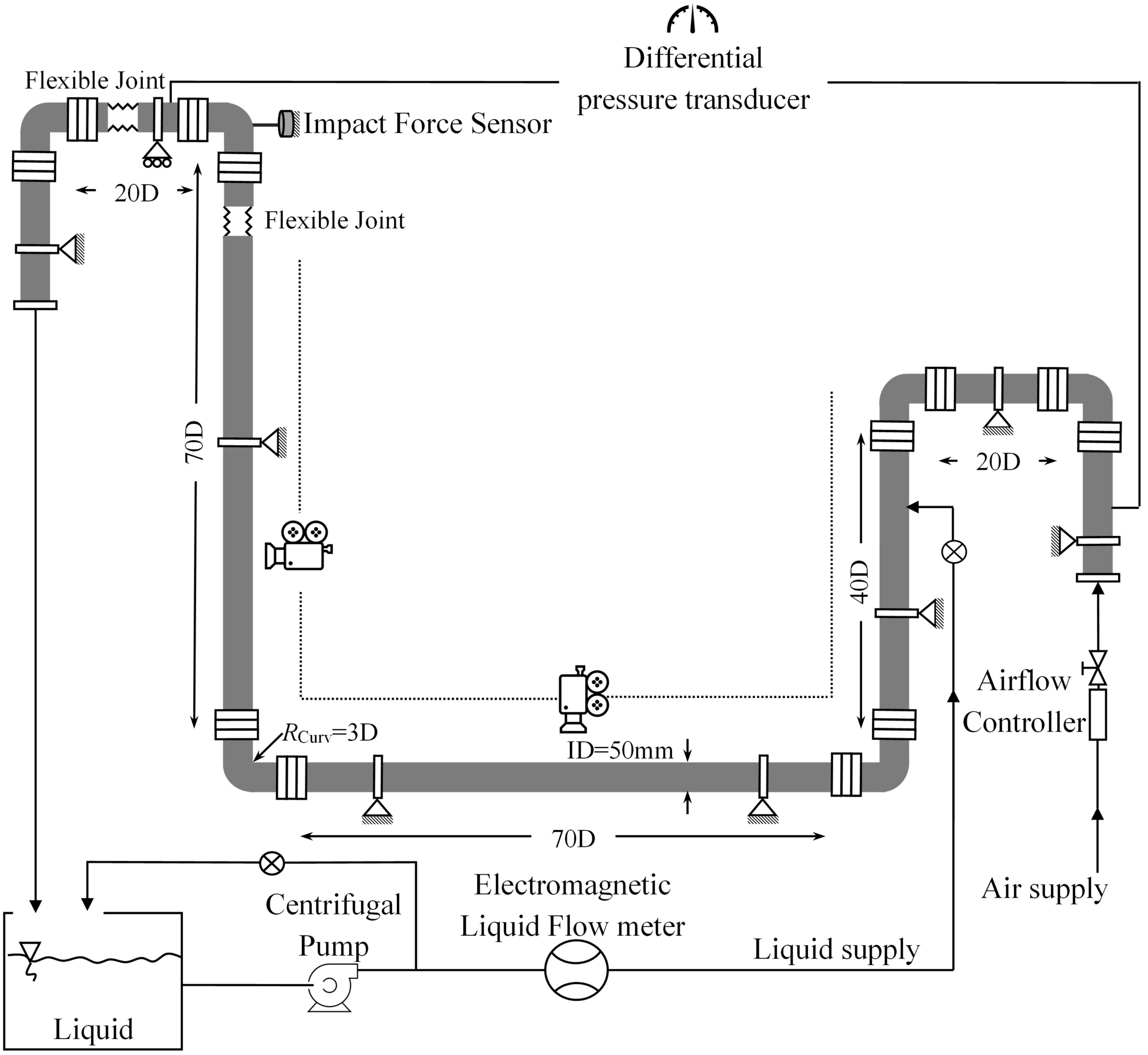

**Figure 2.1**: Schematic description of the air-water experimental setup.

## 3 Numerical model

Two- and three-dimensional (2D and 3D) numerical models were used to predict the transient and turbulent two-phase (gas-liquid) phenomena during the displacement of the liquid and its purge out from the jumper. The tools available in OpenFOAM, an open-source CFD package, were used to obtain the numerical solution. The 2D and 3D numerical model geometries are similar to the experimental system (i.e., in 2D simulation, the lateral distance is equal to pipe ID=50mm).

The boundary conditions imposed are (1) constant or time-dependent mass flow rate of the air at the *inlet*, (2) no-slip condition on the domain *walls,* and (3) constant atmospheric pressure at the *outlet*. At $t$=0 (initial condition), a motionless predefined amount of liquid rests in the low horizontal section of the jumper under gravitational force. Also, at $t$=0, the air mass flow in the inlet is zero. At $t$>0, in all the tested cases, the air mass flow at the inlet changes abruptly (step function) or gradually (ramp-up function) from zero to a constant mass flow corresponding to a specified air superficial velocity at STP ($U^0_{GS}$). The air velocity at the inlet is uniform. Note that the air mass flow should be applied (and not volumetric flow rate or $U^0_{GS}$) since the compressibility of the air cannot be neglected. The ideal gas equation of state ($\rho=P/RT$) was used to calculate the variation of the air density with pressure and temperature. Note that the pressure at the outlet of jumpers rises during the start-up operation due to the developing flow in the downstream pipe (particularly when the downstream pipe is long). However, the outlet pressure rise is relatively slow compared to the simulation time needed to follow the process of water purging from the jumper. Thus, for convenience, the outlet pressure is set constant during the simulation time.

The contact angle was set to a value of $\theta$ = 30°, which represents the measured value of the receding contact angle. The latter was found in our previous studies to affect the downstream propagation of the liquid film tail Bissor et al., 2020; Yurishchev et al., 2023 and thereby the removal of residual liquid amounts from the system.

The "compressibleInterFoam" solver was applied to simulate the flow phenomena. The solver is appropriate for two compressible, non-isothermal, immiscible fluids and uses a VOF (Volume of Fluid) phase-fraction based on the interface capturing approach, Hirt and Nichols, 1981; Roenby et al., 2017. The governing equations are the continuity, momentum, volume fraction and energy equations. With the VOF method, the formulation of the conservation equations considers single-phase flow with the local fluid properties (volume fraction average of the fluids' properties is taken in the interface region cells). The phase fraction is determined as 1 in elements with solely one phase (e.g., water) and as 0 in elements with the second phase (e.g., air). The rest of the cells (with values between 0 to 1) define the interface between the phases. The location (and the thickness) of the interface is obtained by solving the volume fraction equation. Tracking the boundaries of smaller phase domains requires a finer grid and a sharper interface. To enhance the sharpness of the interface, a modification was made to the volume fraction equation by adjusting the "compression coefficient", $C_\alpha$, to a value of 2, as opposed to the default value of 1. Magnini et al., 2018 described the significance of this modification as $C_\alpha = 2$ allows a coarser grid for obtaining converging results, and this value was also used in the current study. It is worth noting that we do not intend to resolve domains of fine dispersion of water in air but to simulate transient flow phenomena at larger scales which are responsible for the flow characteristics of interest that were measured. Finally, RANS turbulence modeling ($\kappa$-$\varepsilon$ or $\kappa$-$\omega$ SST in this study) requires extra transport equations for $\kappa$ (turbulence kinetic energy), and for $\varepsilon$ (turbulence dissipation rate) or for $\omega$ (specific turbulence dissipation rate).

The 2D/3D momentum equations were solved via the PISO velocity-pressure coupling algorithm, by implicit prediction of the velocity field, and explicit first-order pressure correction. Time derivatives were approximated by the first-order Euler scheme to improve solution stability. Gauss linear (second order) scheme was implemented for gradients. Gauss vanLeer interfaceCompression (second order) scheme was used for the convective terms in the volume fraction equation. An interface-compression coefficient is applied to reduce interface smearing as explained above. The convection terms in other transport equations were discretized by Gauss linear scheme. Finally, the interpolation scheme (which finds the value of the variable at a face, given the values at the cells' centroids on either side of that face) is linear. Discretization errors of simulation results (reported and discussed in **Section 5**) were estimated and presented in **Appendix B.**

An adjustable time step was used to ensure the solution stability by limiting the Courant number to 0.5. Time independence tests indicated that decreasing the maximum Courant number to 0.25 does not enhance the accuracy of the solution. However, increasing it up to 1 often results in failure of the solver, when the "negative temperature" error was reported, indicating divergence of the solution. The convergence of the numerical solution was monitored by the residual values of $10^{-6}$ for the velocity, $10^{-7}$ for the pressure. The residuals of $10^{-5}$ were sufficient for the temperature, turbulent kinetic energy ($\kappa$), turbulence dissipation rate ($\varepsilon$) and the specific turbulence dissipation rate ($\omega$). Finally, the phase fraction was satisfied by the residuals of $10^{-8}$. Also, no mass imbalances were detected. The solution was accelerated by parallel computing on 24-64 processors.

It was revealed that solving the numerical problem assuming both incompressible air and liquid (applying "interFOAM" solver similarly to Bissor et al., 2020; Hamami Bissor (Abir) et al., 2021; Kurbanaliev et al., 2019; Larsen et al., 2019; Polansky and Schmelter, 2022; Verma et al., 2017; Yurishchev et al., 2022) is physically wrong since extreme and unrealistically high pressures were predicted for some cases at initial flowtimes. For instance, when the prefilled liquid is completely blocking the air passage in the lowest horizontal part, the pressure may rise to ~1800 atm. However, if the air obeys an equation of state (e.g., ideal gas $\rho = P/RT$), which allows density changes as a function of pressure (and temperature), the predicted pressure variations appear to be reasonable. Other equations of state (e.g., real gases) were not tested since the ideal gas equation for air is valid for the observed pressure ranges (compressibility factor $z$=0.9997÷0.9983 for $p$=1÷5 atm (e.g., Jones, 1983). Note that taking compressibility effects into consideration obviously complicates the numerical solution since an additional (energy) equation must be solved.

A set of tests was performed for the selection of a proper RANS turbulence model, comparing $\kappa$-$\omega$ SST model with two $\kappa$-$\varepsilon$ models (standard and Realizable). The turbulence models, $\kappa$-$\varepsilon$ and $\kappa$-$\omega$ SST, use different approximations, mainly in the near-wall treatment. Therefore, different meshing is required for each model in order to avoid false predictions. The dimensionless normal distance from the wall, $y^+$ of the grid adjacent to the wall is not the same in the two models. The $\kappa$-$\varepsilon$ model recommends $y^+$>30 (i.e., the first grid point should be placed in the log velocity profile region), while the $k$-$\omega$ SST model restricts it to $y^+$<5 (i.e., the first grid point should be placed in the viscous sublayer). The goal of the present two-phase flow simulation is to provide the ability to predict the flow phenomena in both phases (gas and liquid). Assuming a relatively thin liquid layer, an appropriate grid can apparently be structured by strictly obeying the mentioned restrictions for single-gas flow. However, most of the tested cases involve an initially large near-wall region occupied by water. Therefore, the validity of the construction of the grid relying on single-phase gas flow was investigated. The related calculations indicated that the difference between the $y^+$ values (i.e., the difference between the required thickness first mesh cell adjacent to the wall to meet the desired $y^+$ values) corresponding to single-phase flow of the air or water could be up to one order of magnitude, depending on the superficial velocity of the phase. But the water superficial velocity (at the initial stages of the simulation) is significantly lower than the air superficial velocity, thus the actual deviation is lower. As stated previously, the $\kappa$-$\varepsilon$ model has a recommended *minimum* value for cell thickness, so the grid for $\kappa$-$\varepsilon$ should be constructed by considering single-phase air flow, which should be, however, applicable for single-phase water flow as well.

In contrast to the $\kappa$-$\varepsilon$ models, the $k$-$\omega$ SST model does not have a minimum restriction regarding the thickness of the cells adjacent to the wall, meaning that the grid should be constructed by considering liquid single-phase flow. However, a finer grid leads to much more demanding computation. Consequently, it was decided to compromise on a grid for the *k-ω* SST that will be finer than the recommended for gas single-phase flow and slightly coarser than that recommended for liquid flow.

Note that the first cell thickness calculations assume a fully developed turbulent flow, which, however, is not the case in the current study. Initially, the two phases are resting, and at $t$=0 the gas enters the domain, provoking the liquid motion and resulting in a transient flow. Thus, the ability of the proposed non-dynamic grids (especially of *κ-ε*) to detect the majority of the transient phenomena is questionable. It was verified that the grids used for the *κ-ε* and *k-ω* SST models yield grid and time step independent results with a maximum Courant number of 0.5, see **Appendix A**.

The selected 2D grid for computations with *κ-ω* SST turbulence model contained 897,400 cells. The domain face was divided into 70 divisions (in the lateral direction) with a bias growth rate of 1.1 towards both walls. The thickness of the cells adjacent to the domain top and bottom walls obeys the recommended dimensionless normal distance from the wall $y^+$ for the *κ-ω* SST turbulence model for all tested air mass flows (i.e., $y^+$<5). The maximum cell size was limited to 1mm, which allowed limiting the maximum aspect ratio of the cells to an acceptable value of ~10, with a negligible number of cells of a larger aspect ratio. Most of the cells are structured with an aspect ratio<5.

The selected 2D grid (for calculations with *κ-ε* models) contained 90,125 elements. The lateral direction is divided into 25 nonuniform elements with a bias growth rate of 1.1 toward both walls. Maximum axial element length was limited to 3.5 mm, with an allowed maximum cell aspect ratio of <5. Finally, the selected 3D grid (for calculations with *κ-ε* models) contained of 3,071,250 elements. The pipe cross-section included 1125 elements, while the axial element length was limited to 5 mm, so most of the cells obeyed the aspect ratio of <10 (a summary of the elements' distribution for the selected grids is shown in **Table 3.1)**. The 3D domain with *κ-ω* SST was not simulated since the large number of elements of the recommended grid (>20,000,000) required an unrealistic computational effort. It is worth noting that an attempt to solve with *κ-ω* SST on a coarser grid ($y^+$>>5) resulted in false solutions (i.e., the generated numerical solutions do not agree with the observed flow phenomena and experimental data).

**Table 3.1:** Lateral, axial, and total element count for the selected grids.

| | ***Lateral Elements*** | ***Axial Elements*** | ***Total Elements*** |
|---|---|---|---|
| **2D *κ-ω* SST** | 70 | 12820 | 897,400 |
| **2D *κ-ε*** | 25 | 3605 | 90,125 |
| **3D *κ-ε*** | 1125 | 2730 | 3,071,250 |

# 4 Mechanistic modeling

Both the experimental observations and numerical simulation results showed that the air pressure rises during the transient process of displacement and purging out of the accumulated water. The pressure rise is important in determining the loads applied to the pipeline structure. A mechanistic model is herein proposed to enable the prediction of the corresponding pressure rise under various operation parameters (i.e., air mass flow rate, accumulated liquid amount, fluids' physical properties, and jumper geometry). A basic assumption of the model is that the air does not penetrate the liquid phase, and the incompressible liquid moves as a plug. The validity of this assumption is assessed and discussed by comparing the model predictions with experiments (see **Section 5**).

The initial state considered is a hydrostatic equilibrium of the liquid in the jumper under zero mass flow rate of the gas. The liquid level at this stage is $y=L_s$ (see **Figure 4.1**). At $t$=0, the gas is introduced at the inlet through Point 1 at a constant mass flow rate $\dot{m}_{GI}$. The pressure, which is assumed to be uniform between the inlet and the gas-liquid interface (negligible pressure drop in the gas), $P_G(t)$, rises to overcome the hydrostatic and frictional pressure and to accelerate the liquid to velocity $U_L(t)$. In the model, the horizontal outlet section was extended to an infinite pipe in order to mimic natural gas pipeline conditions, where the jumper's outlet is connected to a long pipeline. This is in line with the main goal of the proposed model is to estimate the transient pressure rise during the initial stage of liquid displacement in the jumper, while the late stage of pressure decline associated with the complete purge out of the liquid from the jumper is out of the model scope.

The model enables determining: (1) the liquid displacement in the riser, $H(t)$, (2) the liquid velocity, $U_L(t)$; and (3) the gas pressure, $P_G(t)$, by solving a system of three differential equations. These are derived from (1) velocity continuity at the boundary of the liquid tail, **Eq. [1]**; (2) momentum balance on the moving liquid plug, **Eq. [3]**; and (3) mass balance on the gas occupying the volume trapped between the domain inlet and the moving boundary at the liquid tail, **Eq. [8]**.

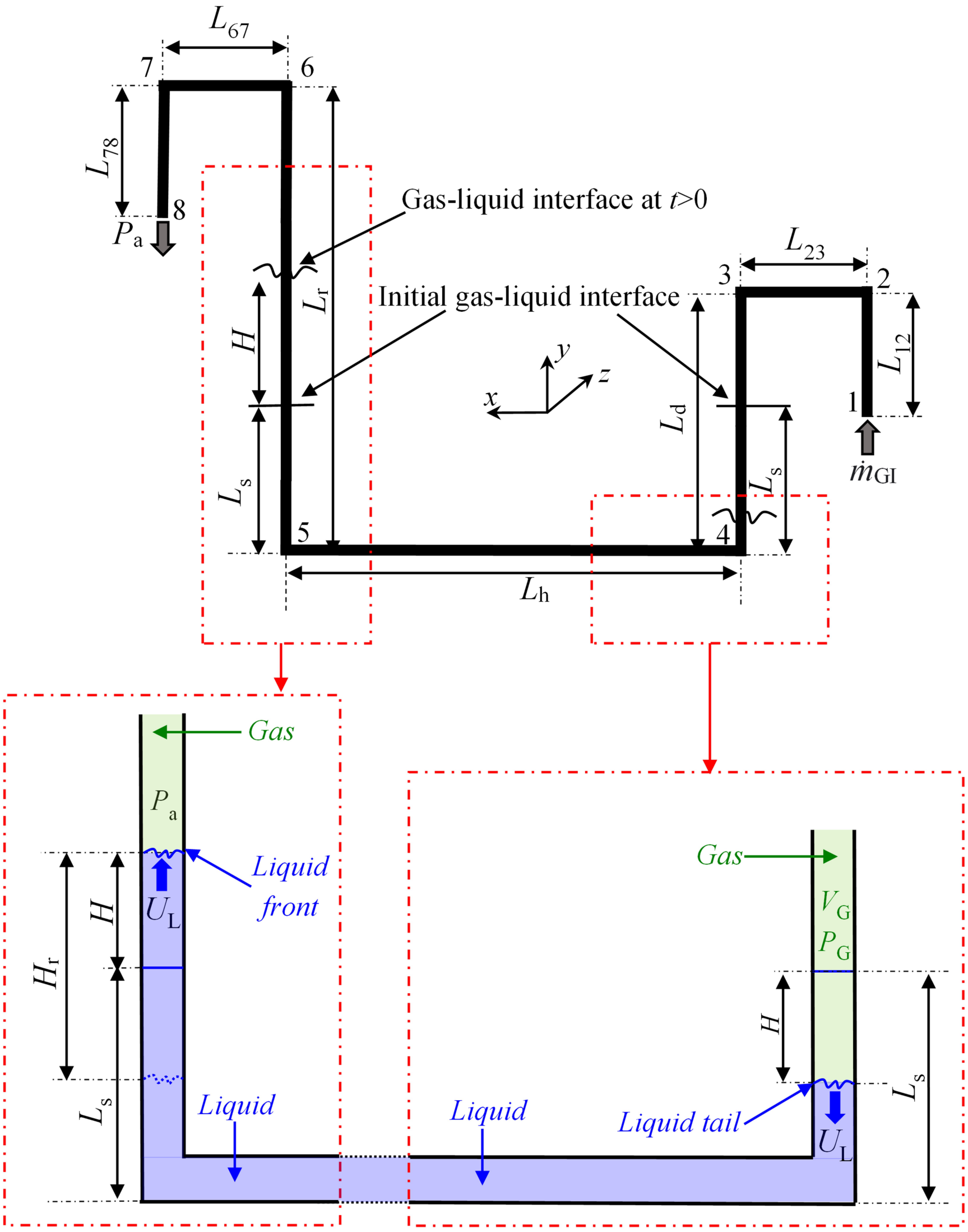

**Figure 4.1:** Scheme of the jumper geometry and gas-liquid dynamics considered in the mechanistic model.

- **Velocity continuity and momentum balance on the moving boundary at the liquid tail**

Continuity of the velocity at the moving interface between the gas and the liquid at the tail yields:

$$\begin{cases} \dfrac{dL_{\mathrm{G}}}{dt} = \dfrac{d\left(L_{\mathrm{G},t=0} + H\right)}{dt} = \dfrac{dH}{dt} = U_L = U_G \\ H(t) = \displaystyle\int_0^t U_{\mathrm{L}} dt \quad ; H_{t=0} = 0 \end{cases} \qquad [1]$$

where $U_{\mathrm{G}}$ and $U_{\mathrm{L}}$ are the gas and liquid velocities at the liquid tail, $L_{\mathrm{G}}$ is the gas domain length from the inlet to the liquid tail, and $H$ is the liquid displacement (height relative to the initial position).

Applying a momentum balance on the liquid body results with:

$$A(P_{\mathrm{G}} - P_{\mathrm{a}}) - AH_{\mathrm{r}}\rho_{\mathrm{L}}g - A\Delta P_{\mathrm{f}} = m_{\mathrm{L}}\frac{dU_{\mathrm{L}}}{dt} = \rho_{\mathrm{L}}AL_{\mathrm{L}}\frac{dU_{\mathrm{L}}}{dt} \qquad [2]$$

Rearranging **Eq. [2]** yields:

$$\frac{dU_{\mathrm{L}}}{dt}=\frac{P_{\mathrm{G}}-P_{\mathrm{a}}-\Delta P_{\mathrm{f}}-H_{\mathrm{r}}\rho_{\mathrm{L}}g}{\rho_{\mathrm{L}}L_{\mathrm{L}}};\quad U_{\mathrm{L}_{t=0}}=0 \tag{3}$$

where $A$ is the pipe cross-section area, $P_G$ is gas pressure at the liquid tail region, $P_a$ is the pressure at the liquid front. $H_r$ is the hydrostatic head due to the difference between the liquid levels in the left (downcomer) and the right (riser) vertical pipes, $m_L$, $\rho_L$, and $L_L$ are the liquid lump mass, density, and length, respectively, and $g$ is the gravitational acceleration. $\Delta P_f$ is the liquid frictional pressure drop calculated based on the wall shear $\tau_f$ obtained by applying the Blasius correlation for the friction factor, $f_L$:

$$\Delta P_{\mathrm{f}}=\frac{4L_{\mathrm{L}}}{D}\,\tau_{\mathrm{f}};\qquad \tau_{\mathrm{f}}=\frac{1}{2}f_{\mathrm{L}}\rho_{\mathrm{L}}U_{\mathrm{L}}^{2} \tag{4}$$

$$f_{\mathrm{L}}=\begin{cases}0.046\mathrm{Re}_{\mathrm{L}}^{-0.2}\rightarrow \text{Turbulent flow}\\ 16\mathrm{Re}_{\mathrm{L}}^{-1}\rightarrow \text{Laminar flow}\end{cases} \tag{5}$$

where $\mathrm{Re}_L=\rho_L U_L D/\mu$

Ignoring the gas frictional pressure gradient ahead of the moving liquid front, $P_a$ is the (atmospheric) pressure at the outlet from the jumper.

- **Hydrostatic head, $H_r$**

The hydrostatic head, $H_r$, varies with time due to the liquid displacement and depends on the instantaneous liquid level in the downcomer and the riser. To evaluate the hydrostatic head, it is convenient to divide the liquid purge-out process into two consequent stages: (1) prior and (2) after the time the front of the liquid reaches the riser top (Point 6 in **Figure 4.1**).

1. Three scenarios can occur before the front of the liquid reaches the riser top, namely when $L_s+H<L_r$
   - $H<L_s$ - The tail of the liquid has not reached the lowest point in the downcomer (Point 4 in **Figure 4.1**). In this case, the hydrostatic head is equal to $H_r=2H$. This is because the front of the liquid moves upwards in the riser, and simultaneously, the tail of the liquid in the downcomer is displaced the same distance downward.
   - $H>L_s$ - The tail of the liquid has reached the lowest point in the downcomer, where the hydrostatic head is equal to the liquid level in the riser $H_r=H+L_s$.
   - The tail of the liquid is already in the riser, although its front has not yet reached the riser top. This can happen when $L_r>L_L$, in which case $H_r=L_L$.
2. Two scenarios can occur after the front of the liquid has reached the riser top ($L_s+H>L_r$). These depend on the liquid initial length in the domain, $L_L$ $(=2L_s+L_h)$, and the riser height, $L_r$:

- The tail of the liquid has not yet left the lowest point of the riser (Point 5 in **Figure 4.1**), since ($L_s$+$H$<$L_L$). In this case, the hydrostatic head is equal to the height of the riser, i.e., $H_r$=$L_r$.
- The tail of the liquid has left the lowest riser point (5) and $L_s$+$H$>$L_L$. Then, $H_r$=$L_r$+$L_L$-($L_s$+$H$).

3. When the liquid tail reaches the riser top (Point 6 in **Figure 4.1**), the whole liquid domain moves in the horizontal section downstream of the riser, whereby $H_r$=0.

- **Mass balance on the compressible gas occupying the volume $V_G$**

A mass balance on the gas occupying the volume, $V_G$, between the inlet and the liquid tail:

$$\frac{dm_G}{dt} = \dot{m}_{GI} \quad [6]$$

Where $\dot{m}_{GI}$ is the mass flow rate of the gas at the inlet. Assuming ideal gas behavior, the mass of the gas upstream the liquid tail is given by:

$$m_G = \frac{V_G P_G}{RT} = \frac{L_G A P_G}{RT} \quad [7]$$

where $T$, and $P_G$ are the gas temperature and pressure, respectively and $R$ is the gas constant. Combining **Eqs. [6, 7]** and **Eq. [1]** and isolating the gas pressure variation with time, assuming constant temperature, yields:

$$\frac{dP_G}{dt} = \frac{RT\dot{m}_{GI} - P_G U_L A}{L_G A}; \; P_{G,t=0} = P_a \quad [8]$$

- **Model modification for low liquid amounts**

In the case of a relatively low accumulated liquid amount (i.e., the initial liquid volume does not entirely occupy the lower horizontal section), the water initially does not block the air passage. Consequently, the model should be slightly modified to preserve the basic assumption of liquid plug flow. To this aim, at $t$=0 the liquid is assembled to form a plug near the lower riser elbow (see **Figure 4.2**). Then, the air is introduced into the domain and removes the liquid (similarly to the above-presented mechanism).

Both experiments and numerical simulations suggest that before the liquid is removed from the domain, it is pushed towards the riser lower elbow forming a pseudo plug that (partially) blocks the air flow.

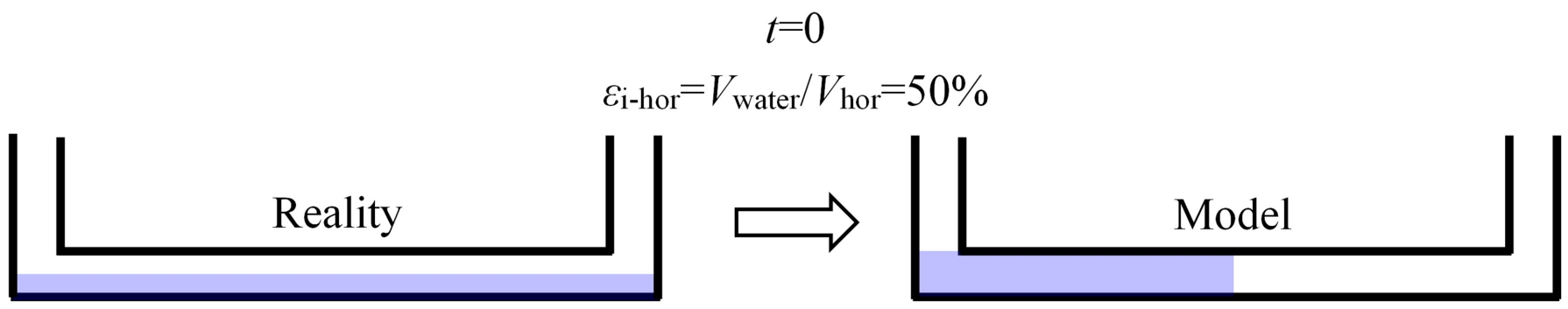

**Figure 4.2:** Illustration of the mechanistic model modification at $t$=0 needed to preserve the liquid plug flow assumption in case of low accumulated water amount

- **Forces on the elbow estimation**

The model predictions of pressure and liquid velocity are also useful for estimating the forces exerted on the jumper elbows due to the liquid passage. In particular, the forces on the riser's upper elbow (Point 6 in **Figure 4.1**) are of interest since its horizontal component produces a moment about the riser's lower elbow (Point 5 in **Figure 4.1**). Also, this elbow receives the highest transient forces (compared to other regions) since the accelerated liquid gains high velocity upon reaching it and the liquid plug still maintains its integrity. The forces acting on the elbow can be deduced from a momentum balance on the fluid flowing through the elbow:

$$\begin{cases} \dfrac{\partial}{\partial t}\iiint \rho u_x dV_{el} + \iint \rho u_x^2 dA_c = -F_x - \iint (p_{out} - P_a) dA_c + \iint (\vec{i} \cdot \bar{\bar{\tau}} \cdot \vec{n}) dA_s \\ \dfrac{\partial}{\partial t}\iiint \rho u_y dV_{el} - \iint \rho u_y^2 dA_c = -F_y + \iint (p_{in} - P_a) dA_c + \iint (\vec{j} \cdot \bar{\bar{\tau}} \cdot \vec{n}) dA_s - \iiint \rho g dV_{el} \end{cases} \qquad [9]$$

where $F_x$, $F_y$, are the force components applied on the riser upper elbow (the force applied on the liquid are $-F_x$, $-F_y$, see **Figure 4.3**); $\rho$, $u_x$, $u_y$ are the local and instantaneous fluid density and velocity components, respectively; $p_{in}$, $p_{out}$ are the local and instantaneous pressures at the inlet and the outlet of the elbow, respectively; $\bar{\bar{\tau}}$ is the viscous stress tensor; $V_{el}$ is the elbow volume; $A_c$, $A_s$ are the flow cross-sectional area at its inlet/outlet and the elbow wetted area, respectively.

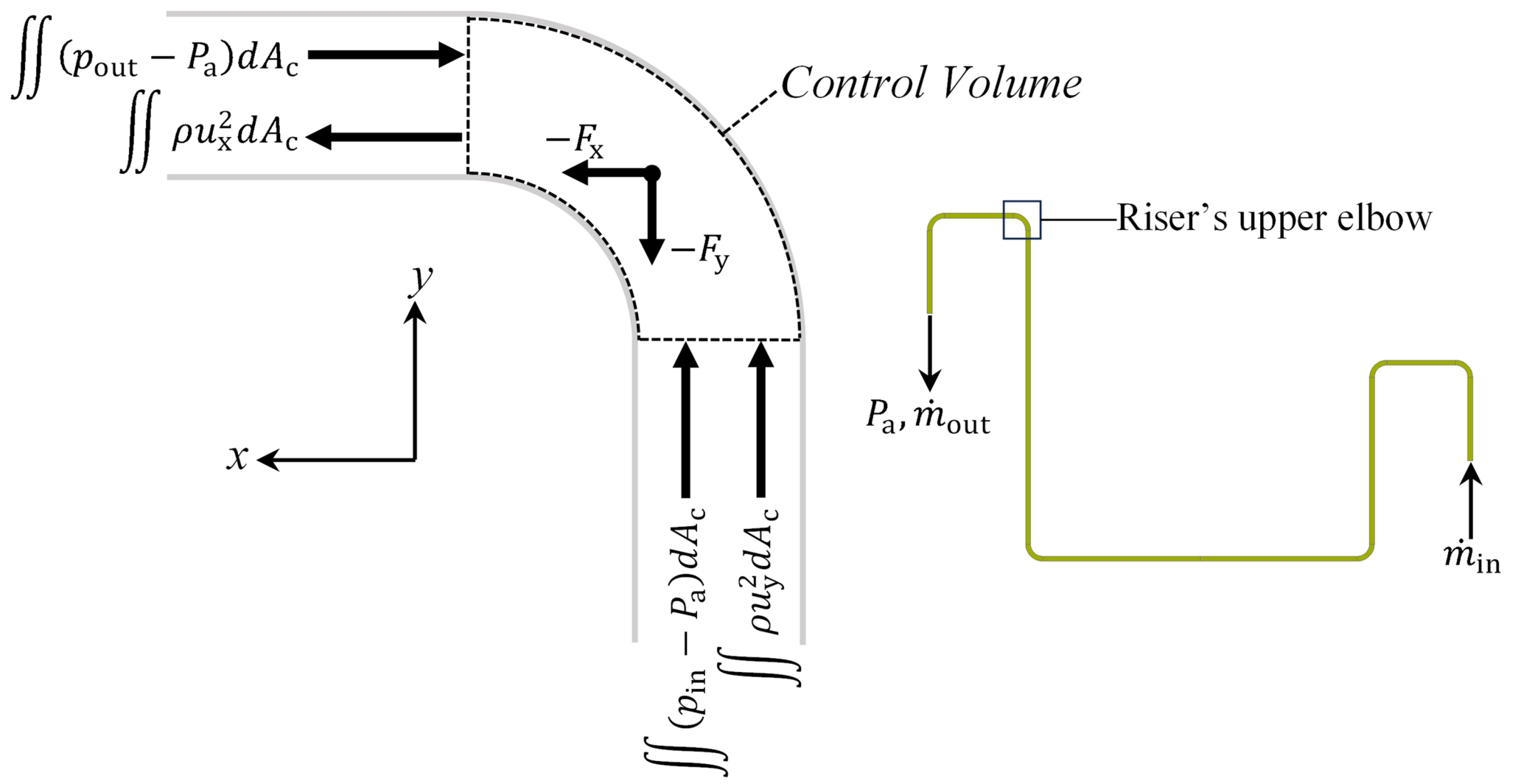

**Figure 4.3:** Free body diagram presenting the forces acting on the riser's upper elbow.

To estimate $F_x$ and $F_y$ via the mechanistic model, a momentum balance is performed on the liquid in the elbow, ignoring the gas flow contribution and the wall friction. Also, the volume of the elbow is considered to be small, so that gravity (body) force and the time required for the liquid plug front (or tail) to pass the elbow are negligible. Accordingly, the time variation of the liquid momentum while passing the elbow is also ignored in the momentum balance. Under these assumptions, $F_x = F_y$.

The time variation of $F_{x|y}$ is then obtained by considering the momentum balance on the liquid plug during the following stages of its displacement:

(*a*): $H<L_r$ – the liquid plug front has not yet reached the riser's upper elbow (no liquid in the elbow and its downstream section), hence $P_6 \sim P_a$ ($P_6$ is the pressure at the riser's upper elbow, point 6 in **Figure 4.1**), whereby:

$$F_{x|y} = 0 \qquad [10]$$

(*b*): $L_r < H < L_r + L_s$ – the liquid plug passes through the riser's upper elbow (its front is located downstream to the elbow, but its tail is located upstream of the elbow):

$$-F_{x|y} = (P_6 - P_a)A + \rho_L U_L^2 A \qquad [11]$$

where $U_L$ is the liquid velocity while passing the elbow during the transient process (calculated by **Eq. [3]**). At this stage, the force results from the pressure in the elbow and the liquid inertia components. It is worth noting that $U_L < U^0_{GS}$ upon reaching the upper elbow. For instance, in the case of $\varepsilon_{i\text{-}hor}$=50%, $t_{ramp}$=2s: the liquid velocity, $U_L$, upon reaching the upper elbow is 7.43, 8.63 and 9.52 m/s for $U^0_{GS}$=10, 15, and 20 m/s, respectively.

(*c*): $H>L_r+L_s$ – the liquid plug tail has left the riser's upper elbow (no liquid in the elbow), so the force is $(P_6\text{-}P_a)A$.

# 5 Results and discussion

We report and discuss the experimental data, numerical simulations and mechanistic models' predictions on the flow phenomena taking place in the riser and its lower elbow during the liquid displacement by the gas flow. We will focus on the critical gas velocity required for purging the accumulated liquid. The pressure drop and forces acting on the jumper's structure will also be addressed.

## 5.1 Flow characteristics

The experiments and numerical simulations conducted are aimed at examining the flow phenomena taking place in the jumper upon the gas production restart (following shutdown) when some liquid has accumulated in its low horizontal part. The flow characteristics during the transient process of liquid displacement and purge out depend on the gas flow rate and the initial liquid amount. In the following, the critical gas velocity is defined as the minimal gas velocity needed to purge completely the initially accumulated liquid from the jumper.

### *5.1.1 Experimental observations*

In the experiments, the gas flow rate was increased to the desired final value following a linear set point ramp-up lasting 2 seconds in all the cases reported in this section. At relatively high (supercritical) air superficial velocities such as $U^0_{GS}$>20 m/s, practically all the liquid is washed out of the experimental facility already in the first stages of the air flow. Only a very small amount of liquid remains in the system in the form of drops hanging on the riser walls or slowly crawling upward toward the exit of the system (see **Figure 5.1.a**). Eventually, the drops evaporate into the air flow, and the pipe dries out completely. It should be noted that water evaporation takes place in our experiments since the air entering the system is not saturated. In a gas pipeline, such evaporation will not occur since the gas is saturated.

Lower (subcritical) gas velocities ($U^0_{GS}$≈18 m/s) also allow fast removal of most of the liquid. However, in this case, annular thin film flow was observed for a relatively long time. From the visual observation, it seemed like the liquid in the annulus was moving upwards, but in fact, no liquid reached the outlet during the simulation time. Later, the film broke and disintegrated into droplets, which eventually evaporated. (see **Figure 5.1.b&c**). At $U^0_{GS}$≈16 m/s (**Figure 5.1.d**), wavy annular flow is clearly observed, with liquid backflow in some regions. Apparently, the annular film thickness is sufficiently high, and even though some evaporation may take place, this flow pattern lasts over a long period of flow time. A further decrease in the gas flow rate leads to a destabilization of the annular flow configuration. Indeed, at $U^0_{GS}$≈12 m/s (see **Figure 5.1.e**) most of the liquid falls and rises periodically, resulting in churn flow. Overall, the flow dynamics appear disordered and random, and water recirculation regions are clearly observed. Finally, at $U^0_{GS}$≈8 m/s, the gas flow rate is so low that the liquid periodically blocks parts in the riser (see **Figure 5.1.f**). During the flow passage blockage, the gas pressure increases, and upward shooting of the liquid plug is observed. However, the liquid inertia is not high enough to overcome the backward gravitational force. Consequently, the liquid falls back and re-blocks the riser's lower elbow.

Note that low gas velocities ($U^0_{GS}$<8 m/s) were also examined. The flow patterns in these cases are dependent not only on the initial (and the residual) liquid amount in the system but also on the gas velocity. Bubbly (or elongated bubble) flow through the water may prevail if the initial liquid amount is extremely high (i.e., it completely blocks the air passage, $\varepsilon_{i\text{-}hor}$>100%) and the gas velocity is sufficiently low ($U^0_{GS}$<4 m/s). For low initial liquid amounts and low gas velocities, most of the liquid remains in the lower horizontal section, and stratified smooth or stratified wavy flow patterns are observed. Here, however, we focus on gas velocities that are sufficiently high to eventually displace most of the liquid from the horizontal section into the riser and result in the flow patterns shown in **Figure 5.1.**

The transparency (Perspex) of the riser's lower elbow enables the documentation of the two-phase air-water dynamics in the elbow. Examining the elbow region is crucial for determining whether a complete liquid purge out from the system is achievable at a given flow rate. For $U^0_{GS}$<14 m/s (see example **Figure 5.2.a**), the water circulates in the elbow region. Increasing the air flow rate leads to a more stable flow pattern, with water lumps residing at the elbow bottom and upper walls (see **Figure 5.2.b**). At higher $U^0_{GS}$(≈>18 m/s), a negligible volume of water (in the form of small droplets or larger chunks) may remain in the elbow. Those droplets eventually evaporate into the air flow or are pushed into the riser and crawl upwards (see **Figure 5.2.c**). Note that the accuracy of the reported experimental critical gas velocity is actually determined by the steps by which the gas flow rate was altered in the experimental runs, i.e. ±1 m per second (rather than the accuracy of the instrument (given in **Table 2.1**). The observations described above indicate that at subcritical air velocities ($U^0_{GS}$<18 m/s), the gas flow is incapable of sustaining the stable annular flow, and other unstable patterns prevail in the elbow region. In fact, if the water backflow in the riser cannot be prevented, it returns to the elbow region (and/or circulates in the riser). Hence, in the experimental setup, the minimal (critical) air velocity ($U^0_{GS|Crit}$) needed to purge from the jumper the entire amount of initially accumulated water is ≈ 20 m/s.

The flow patterns during the transient process of the water displacement and removal from the jumper at critical (and supercritical) gas velocity are dependent on the initial amount of water. In fact, also at subcritical velocities, depending on the initial amount of liquid, a major part of the liquid may be pushed to the top of the riser and out of the jumper shortly after the inception of airflow. Yet, the $U^0_{GS|Crit}$ value, which should be sufficiently high to remove the final small amount of liquid, is independent of the initial amount of water in the jumper.

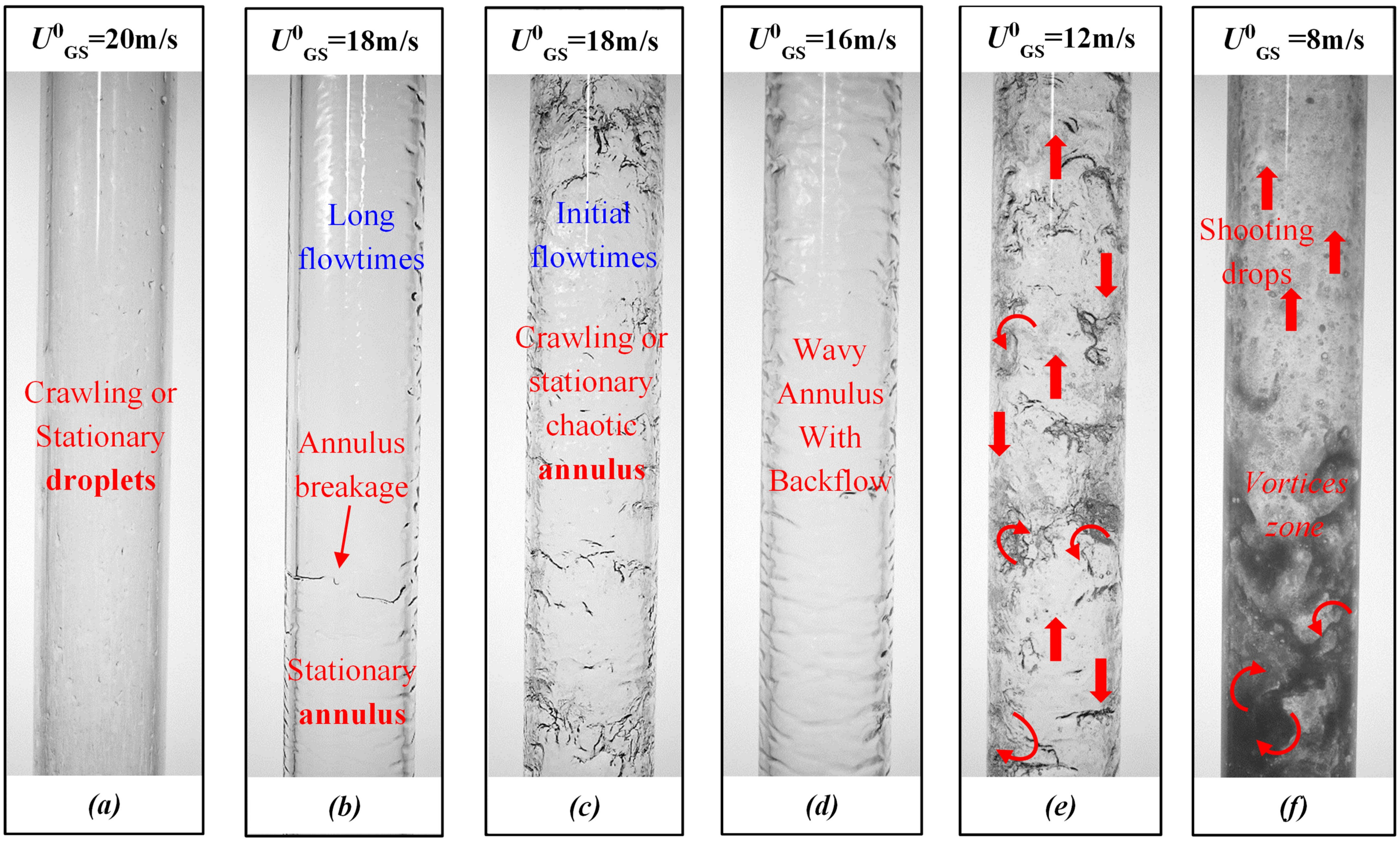

**Figure 5.1:** Observed two-phase flow patterns in the riser (~20D downstream of the riser's bottom elbow) for various gas velocities, $U^0_{GS}$. Initial water volume ~3500 ml (corresponding to $\varepsilon_{i,hor}$=0.5).

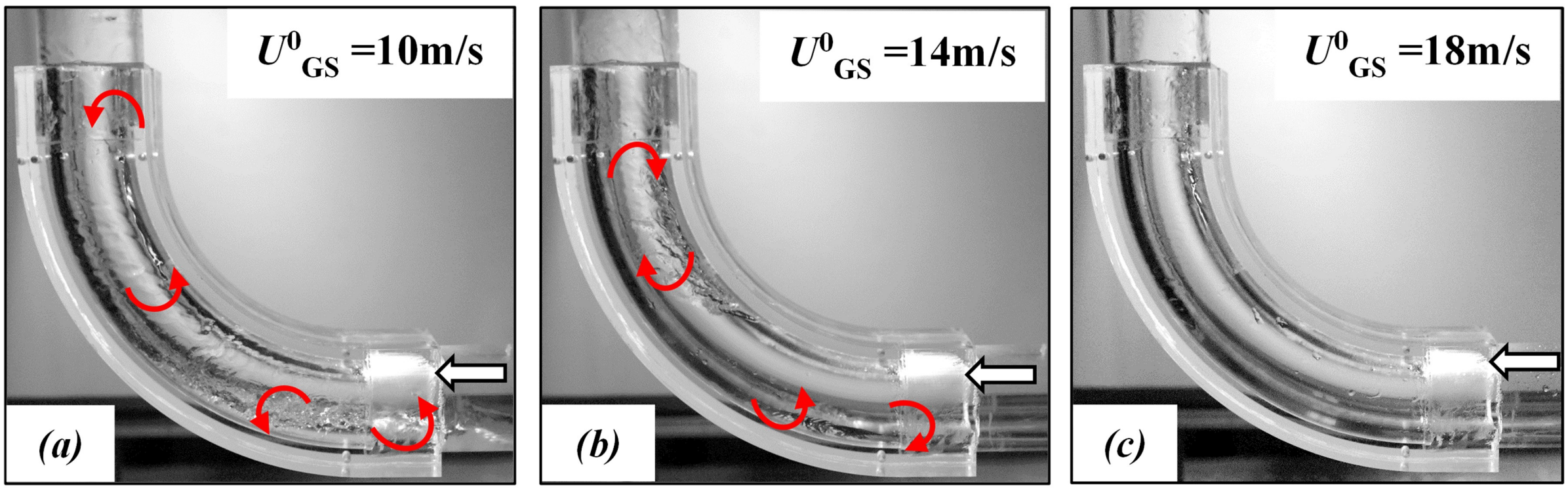

**Figure 5.2**: Two-phase flow observed in the riser's bottom elbow for various gas velocities, $U^0_{GS}$. Initial water volume ~3500 ml ($\varepsilon_{i\text{-}hor}$=0.5).

Wallis, 2020 presented a common criterion for avoiding flow reversal in the continuous film:

$$U_{GS}\sqrt{\frac{\rho_G}{(\rho_L-\rho_G)gD\sin\beta}} \geq C \qquad [12]$$

where $\beta$ is the pipe inclination angle and $C$ is a constant= 0.8÷1

Apparently, this criterion is a relevant tool for predicting the critical gas velocity needed to remove the accumulated liquid from the system through the riser of the M-shaped jumper. Accordingly, for atmospheric air-water flow in the riser, the flow reversal will be avoided when the gas superficial velocity, $U^0_{GS}$, exceeds the following threshold:

$$U^0_{GS|Crit|FRC} = C\sqrt{\frac{(\rho_L-\rho^0_G)gD}{\rho^0_G}} \approx 16 \div 20 \text{ m/s (for } C = 0.8 \div 1) \qquad [13]$$

where $\rho^0{}_G$ is gas pressure at standard pressure and temperature conditions and "FRC" stands for "Flow Reversal Criterion."

Another criterion for predicting the critical gas velocity considers the gas velocity required to break the liquid into non-deformable drops, Brauner, 2003. The critical Weber number for drop breakage can be related to the maximal drop size by Brodkey, 1969 empirical correlation, which is widely used to evaluate $d_{max}$ in pneumatic atomization:

$$\text{We} = \frac{\rho_G U^2_{GS} d_{max}}{\sigma} > 12(1+1.006\text{Oh}^{1.6}) \qquad [14]$$

where $\sigma$ is the surface tension, and Ohnesorge number, Oh=$\mu_L/(\rho_L d_{max}\sigma)^{0.5}$ (can be ignored for low viscosity liquids, e.g., water). The maximal drop size should be smaller than that of a deformable drop, $d_{crit}=[0.4\sigma/(\rho_L-\rho_G)g]^{0.5}$, whereby the critical drop diameter corresponds to a critical Eötvös number, Eo=$(\rho_L-\rho_G)d^2_{crit}/\sigma$=0.4. Hence, according to this criterion, the $U^0_{GS}$, should exceed the following value:

$$U^0_{\mathrm{GS|Crit|GIC}} = \left(\frac{360\sigma\rho_{\mathrm{L}}g}{{\rho^0_{\mathrm{G}}}^2}\right)^{0.25} \approx 20.5\ \mathrm{m/s} \quad [15]$$

where "GIC" stands for "Gas Inertia Criterion".

The criterion for breaking the liquid phase into drops of $d_{\mathrm{max}} < d_{\mathrm{crit}}$ does not guarantee that the produced drops will be elevated and removed from the riser. Actually, the gas drag force should be sufficiently high to avoid settling of the drops due to gravity. The force balance on the maximal (non-deformed) spherical drop reads:

$$\frac{1}{2}C_{\mathrm{D}}\frac{\pi d^2_{\mathrm{max}}}{4}\rho_{\mathrm{G}}\left(U_{\mathrm{GS|Crit|D}}\right)^2 > \frac{\pi d^3_{\mathrm{max}}}{6}(\rho_{\mathrm{L}} - \rho_{\mathrm{G}})g \quad [16]$$

where $C_{\mathrm{D}}$ is the drag coefficient and $U_{\mathrm{GS|Crit|D}}$ is the critical gas velocity based on force balance on a drop. Using **Eq. [14]**, (i.e., $\mathrm{We}_{\mathrm{Crit}} = 12$), **Eq. [16]** yields:

$$U^0_{\mathrm{GS|Crit|D}} > \left(\frac{16\sigma\rho_{\mathrm{L}}g}{C_{\mathrm{D}}{\rho^0_{\mathrm{G}}}^2}\right)^{0.25} \quad [17]$$

Considering the values of $C_{\mathrm{D}}$ experienced by drop in the intermediate and Newtonian regions ($C_{\mathrm{D}}$<1), $C_{\mathrm{D}}^{0.25} \cong 1$, whereby $(U^0_{\mathrm{GS|Crit|GIC}})/(U^0_{\mathrm{GS|Crit|D}})>1$. Hence, once the criterion of $U^0_{\mathrm{GS|Crit|GIC}}$ is satisfied, the gas velocity is already sufficiently high to lift the drops upward ($U^0_{\mathrm{GS|Crit|GIC}} > U^0_{\mathrm{GS|Crit|D}}$). Although an exact comparison is impossible (as the experimental gas velocity was changed by steps of 1 m/s), the critical gas velocity of $U^0_{\mathrm{GS|Crit|Exp}} \approx 20$ m/s obtained in the current experimental setup agrees with the values predicted by both criteria ($U^0_{\mathrm{GS|Crit|FRC}}$ and $U^0_{\mathrm{GS|Crit|GIC}}$). Note that the criterion for $U^0_{\mathrm{GS|Crit|GIC}}$ is similar to the criterion suggested by Turner et al., 1969 and revised by Belfroid et al., 2008. However, the constant in their model (6.7) compared to 4.356 in **Eq. [15]** results in a critical gas velocity of ≈ 30.6 m/s, which largely overpredicts the experimental critical gas velocity.

### *5.1.2 Numerical simulations*

2D and 3D simulation results are shown below for $U^0_{\mathrm{GS}}$ > 10 m/s. Obviously, the complex two-phase flow patterns obtained by the 2D and 3D simulations are not exactly the same, however, the principal flow dynamics predicted are similar. Upon introducing the gas flow into the numerical domain, the liquid is propelled toward the riser's bottom elbow and penetrates the riser section. The manner of the liquid displacement depends on the gas velocity ($U^0_{\mathrm{GS}}$) and the initial amount of liquid accumulated in the jumper (i.e., $\varepsilon_{\mathrm{i\text{-}hor}}$, the initial liquid holdup in the lower horizontal section). A comparison of the observed flow pattern in the horizontal section at short flow times and the simulation results is shown in **Figure 5.3** ($U^0_{\mathrm{GS}}$=16 m/s $\varepsilon_{\mathrm{i\text{-}hor}}$=0.5). The time scale is normalized by the gas residence time ($\tau_{\mathrm{r}}=L/U^0_{\mathrm{GS}}$, where $L$ is the Jumper length (13 m) and $U^0_{\mathrm{GS}}$ is the gas superficial gas velocity at STP). Thus, the normalized time scale is defined by $t/\tau_{\mathrm{r}}$.

Experimental observations indicate that under such conditions, upon introducing the air flow, a liquid plug, which locally blocks the air flow is formed (see **Figure 5.3.g**). The plug is accelerated downstream and becomes highly aerated while being pushed out of the experimental domain. However, some of the tested numerical models were not able to predict the air passage blockage and the following plug acceleration. **Figure 5.3.a** shows that 2D simulations with the $\kappa$-$\omega$ SST turbulence model predict plug formation and air passage blockage relatively far from the down-comer elbow. With the Realizable $\kappa$-$\varepsilon$ model (in both 2D and 3D simulations), a blockage of the air flow was documented near the downcomer elbow, similar to the corresponding experiment. Yet, the flow pattern in the horizontal section predicted by the standard $\kappa$-$\varepsilon$ model (by both 2D and 3D simulations) is completely different from that observed in the experiments (**Figure 5.3**.**c&f1**). No water plug formation is predicted, and stratified flow was preserved in the horizontal section. This is attributed to the fact that the standard $\kappa$-$\varepsilon$ model is a high Re formulation, which is generally appropriate for developed turbulent flows, rather than for simulating transient conditions with initially stationary fluids. Eventually, at longer simulation times, the liquid assembled in the riser's bottom elbow, and only then the air passage is partially blocked (not shown). The flow pattern in the low horizontal section of the jumper was found to be important for understanding the pressure variations during the displacement and flush out of the accumulated water by the airflow (see **Section 5.2**).

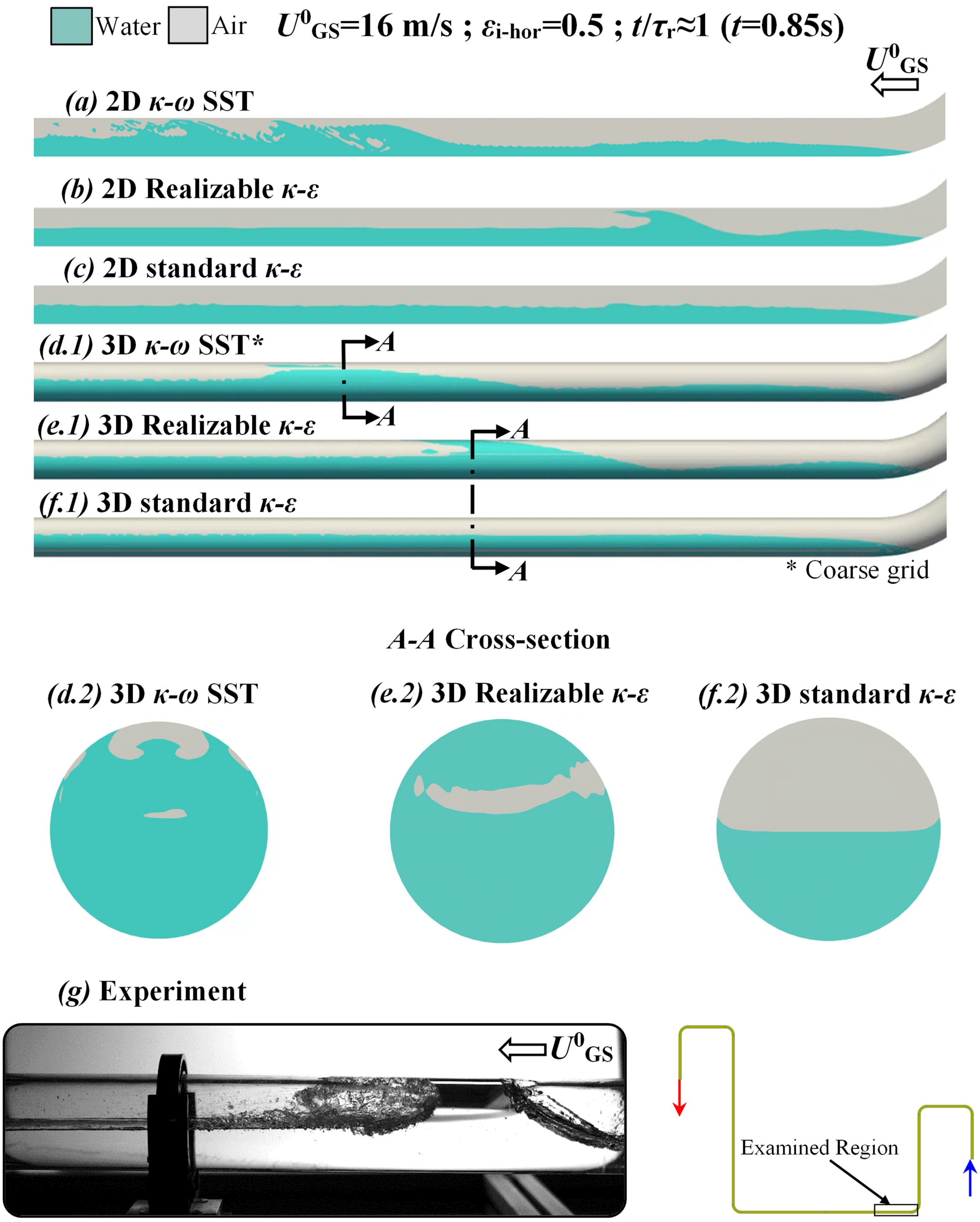

**Figure 5.3***:* Typical locations of the air-water phases at short gas residence times ($t/\tau_r$<1.5) in the low horizontal section downstream of the down-comer elbow. Numerical predictions of 2D models (*a, b,* and *c*), the 3D models (*d, e,* and *f*), and the experimental observation (*g*).

The 3D simulations indicate that if the air flow rate is not high enough to sustain an annular flow in the riser, a water plug that reaches the riser disintegrates into smaller chunks at some distance downstream of the elbow, similarly to the experimental observations (see **Section 5.1.1**). In that case, the water chunks are entrained into the air flow and may break into smaller drops or merge with others. It is often discussed in the literature that the VOF algorithm for tracking the phases interface is not ideal for dealing with such flow patterns (e.g., Černe et al., 2000; Wardle and Weller, 2013a; Yurishchev et al., 2022). Indeed, the simulations showed that with the breakage of the water lumps, the phases interface becomes more diffused than in segregated air-water flow, and the effective density of the chunks is lower than that of water. Therefore, these simulated chunks are lifted by the airflow and eventually, leave the simulation domain more easily than the "real" water chunks. Consequently, in the absence of continuous liquid supply into the jumper, churn flow cannot be simulated for a long flow time by the VOF method. With subcritical $U^0_{GS}$ values, wherein in the experiments, water was clearly observed in the riser, the simulations showed that after a long flowtime, liquid remained only in the elbow region. Therefore, the simulated $U^0_{GS|Crit}$ was determined as the $U^0_{GS}$ value needed to prevent the liquid backflow in the riser. To this aim, the simulated velocity profiles were examined at several cross-sections along the riser. At subcritical $U^0_{GS}$ negative vertical velocity was observed in the water phase, while at supercritical $U^0_{GS}$ both phases' velocities were positive (see example **Figure 5.4**).

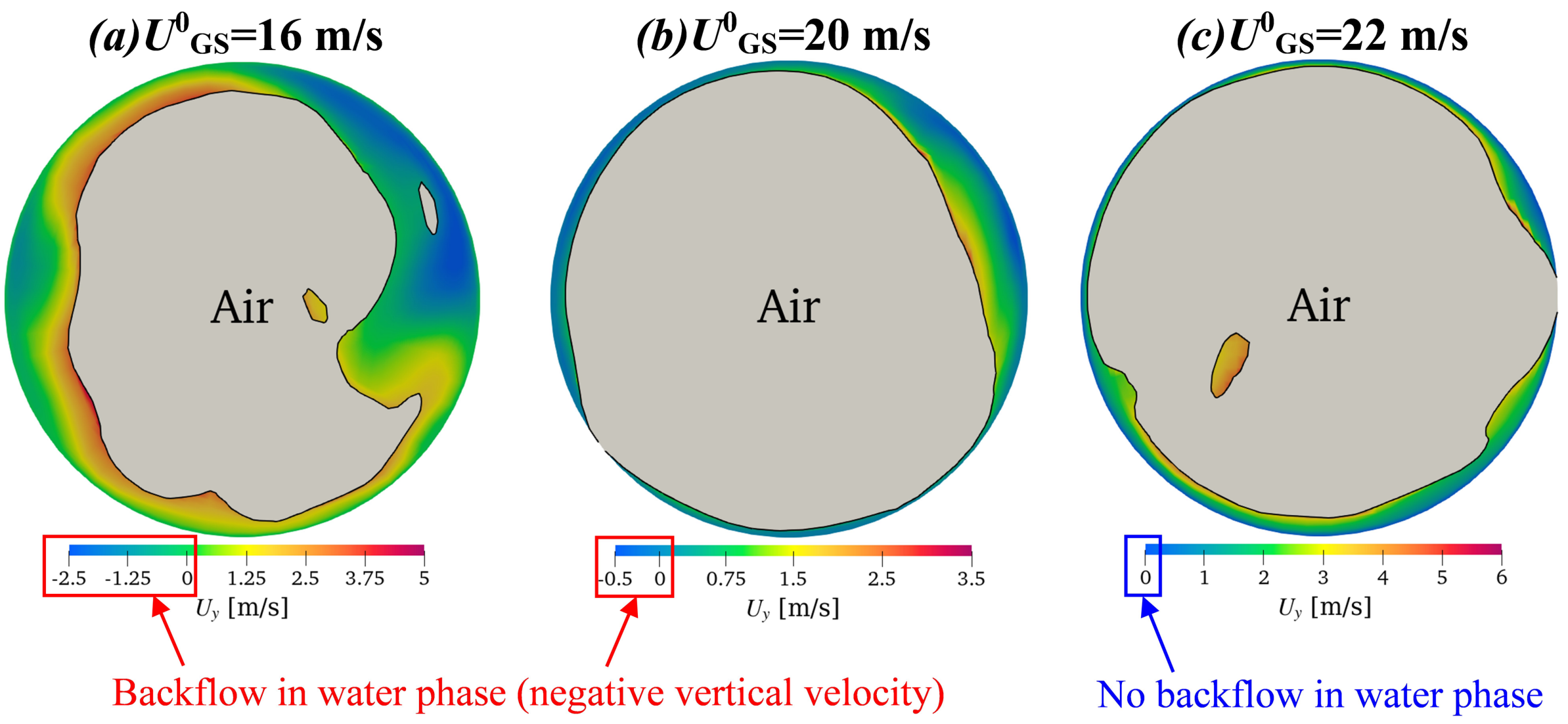

**Figure 5.4***:* Water vertical velocity ($U_y$) field at $t/\tau_r \approx 12$ obtained with the 3D Realizable *κ-ε* numerical model, cross-section at 15D downstream to the low upcomer elbow, (*a,b*) $U^0_{GS}$=16, 20 m/s, subcritical air velocity and (*c*) $U^0_{GS}$=22 m/s supercritical air velocity. Note: the grey area is related to the airflow. The upward in-situ air velocity is much higher than the water velocity.

The air-water hydrodynamics in the bottom elbow of the riser at subcritical $U^0_{GS}$ is also of interest. **Figure 5.5** compares the flow pattern and water amount in the elbow obtained by various combinations of 2D, 3D and turbulence models. In comparison with the flow visualized in the experiments (**Figure 5.2**), the 3D standard *κ-ε* simulations failed to predict the flow in the elbow region, while 2D *κ-ω* SST and 3D Realizable *κ-ε* seem to provide more realistic results.

**Table 5.1** summarizes the minimal gas velocity needed to prevent backflow in the riser obtained by the numerical simulations and the experiments. It was found that 2D *κ-ω* SST and 3D Realizable *κ-ε* simulations provided the closest results to the experimental value. Both the 2D and 3D standard *κ-ε* simulations underpredict the experimental $U^0_{GS|Crit}$ (by ~25%), whereas the 2D Realizable *κ-ε* simulations overpredict the experimental value (by >25%).

**Table 5.1:** Summary of the minimal gas velocity ($U^0_{GS|Crit}$) needed to prevent backflow in the riser (low and high values indicate sub and supercritical gas velocities)

| | 2D | | | 3D | | Experimental |
|---|---|---|---|---|---|---|
| | standard *κ-ε* | Realizable *κ-ε* | *κ-ω* SST | standard *κ-ε* | Realizable *κ-ε* | |
| $U^0_{GS|Crit}$ **[m/s]** | 14-16 | 24-26 | 20-22 | 12-14 | 20-22 | ≈20 |

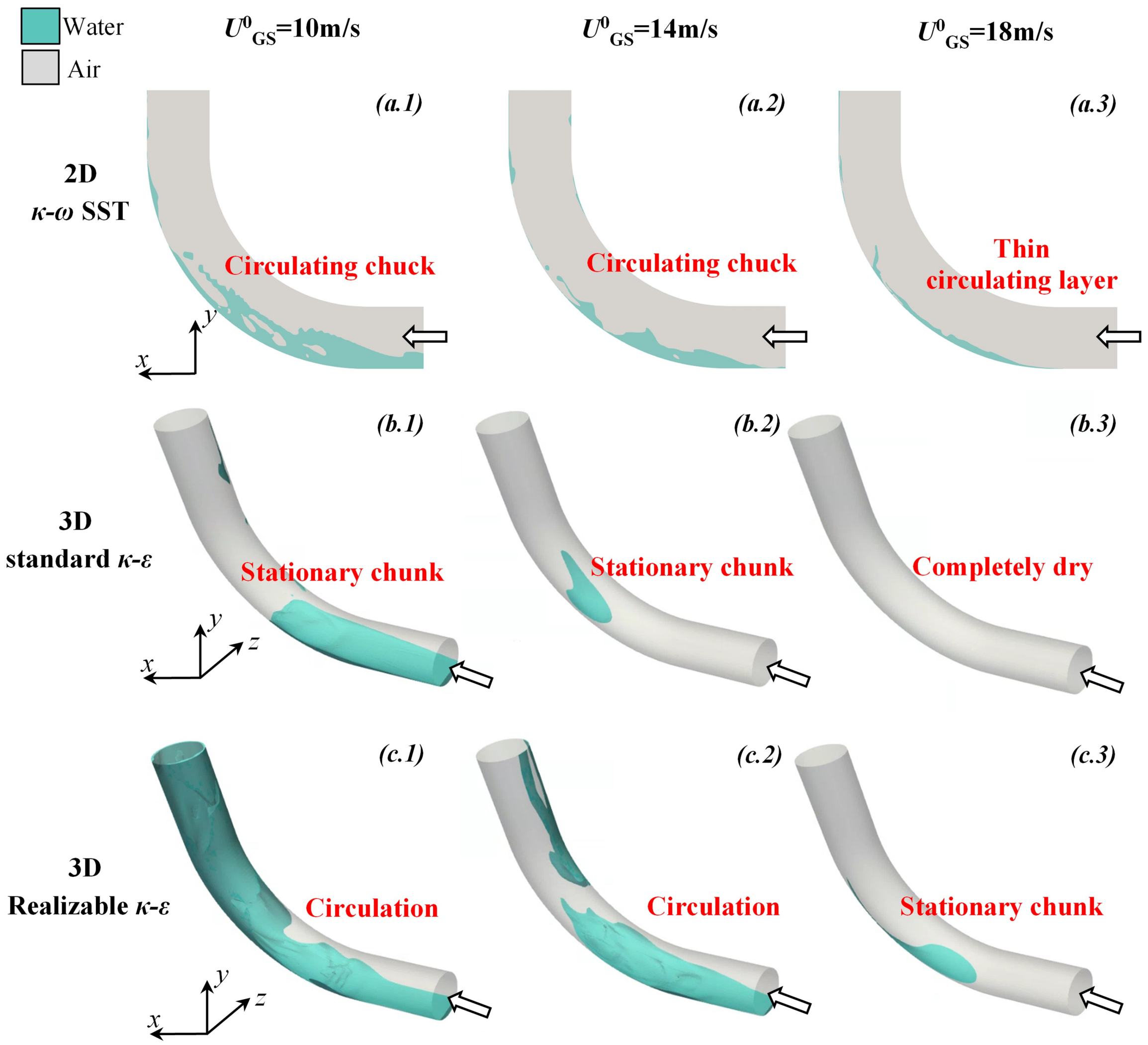

**Figure 5.5**: Comparison between the numerical models (2D (*κ-ω*) vs 3D (standard and Realizable *κ-ε*)) regarding the flow patterns (and liquid residual) in the bottom riser elbow for various air velocities, $U^0_{GS}$, at long flowtime. These stills can be compared to experimental observations presented in **Figure 5.2**.

## 5.2 Pressure response

Pressure response monitoring during a startup protocol may be translated to loads (forces and moments) exerted on the jumper structure (as elaborated in **Section 5.3**). Therefore, it is of interest to introduce reliable tools for its correct prediction. **Figure 5.6** shows some examples of the pressure variations obtained by numerical simulations, mechanistic modeling, and experiments ($\varepsilon_{i\text{-}hor}$=0.5 with ramp-up of 2 seconds between zero to the final gas flow rate (corresponds to $t/\tau_r \approx 2.46$ and $t/\tau_r \approx 3.07$, for $U^0_{GS}$=16 m/s and $U^0_{GS}$=20 m/s, respectively).

Comparing the predictions obtained by the numerical simulations and the mechanistic model to the measurements (see **Figure 5.6**) indicates that both modeling approaches provide reasonable agreement with experimental data of the initial (and maximal) pressure peak value. This can be expected since a major contribution to the initial pressure rise under these conditions is attributed to the liquid acceleration by the gas flow. The pressure peak obtained with the standard $\kappa$-$\varepsilon$ turbulence model is usually lower (and delayed), due to inaccurate prediction of the two-phase dynamics at the initial flow times for relatively low $U^0_{GS}$ (see discussion with reference to **Figure 5.3**). However, in most cases, none of the methods successfully predict the time at which the pressure peak occurs.

As seen in **Figure 5.6**, the initiation of pressure rise predicted by the model is earlier than those predicted by numerical models and the results obtained in the experiment. The reason for this is the different initial conditions and the liquid position in the model. As explained in **Section 4**, at $t$=0, the liquid is assembled as a plug near the riser's bottom elbow. Therefore, the liquid pug displacement and the associated pressure rise start immediately as airflow is introduced to the system. However, in the experiments and the simulations, at $t$=0, the liquid is resting in the horizontal section (under gravitational force), and some delay is expected until the water plug is created (see discussion with reference to **Figure 5.3**). Another possible reason for the different time responses could be the differences between the model and the experiments' airflow ramp-up profile during the ramp-up period. Note that shifting the pressure response in the mechanistic model (~$\Delta t/\tau_r \approx +1$) leads to better agreement between the model, the experimental data, and simulations in terms of pressure rise initiation. However, in the simulations, the predicted pressure rise rate is similar for all tested numerical models (except for both 2D and 3D standard $\kappa$-$\varepsilon$) and is slightly faster compared to experimental data. Nevertheless, the deviation is acceptable (see a detailed discussion in **Section 5.2.1**).

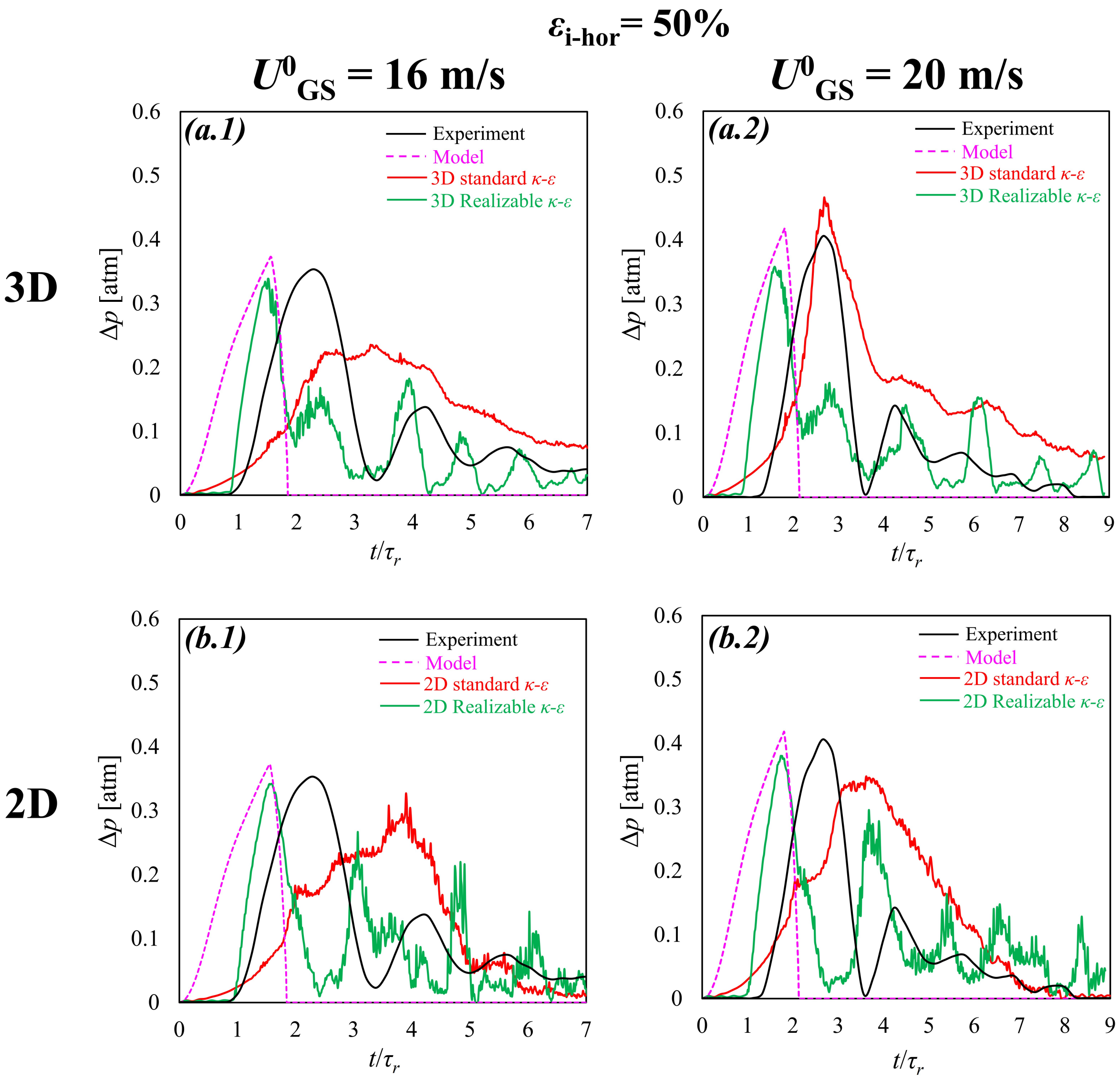

**Figure 5.6:** Variation of the pressure drop over the riser during the purge-out of accumulated water. Conditions: $\varepsilon_{\text{i-hor}}$=0.5 and ramp-up period $t_{\text{ramp}}$=2 s (corresponding to $t/\tau_r$≈2.46 and $t/\tau_r$≈3.07, for $U^0_{\text{GS}}$=16 m/s (LHS figures) and $U^0_{\text{GS}}$=20 m/s (RHS figures), respectively). Comparison of results obtained by the various numerical models used: 3D (*a*) and 2D (*b*) with experimental data (the experimental error is provided in **Table 2.1**) and mechanistic model predictions.

It is notable that during the pressure decline, pressure fluctuations were documented both in the experiments and in the numerical simulations. Apparently, the fluctuations are caused by the formation of liquid plugs/bridges (associated with pressure rise) in the riser followed by their collapse (related to pressure decline). The detected pressure fluctuations decay over several cycles as the water is purged from the system. The high-frequency, low amplitude pressure fluctuations ($f$>10 Hz) documented in the simulations are due to the complex variations in the phases' distributions and are considered to be less significant. The more significant low-frequency (~1 Hz) pressure fluctuations, which were also observed in the experiments, are of interest. The results indicate that both 2D and 3D standard $\kappa$-$\varepsilon$ models usually failed to produce any low-frequency pressure fluctuations, while in the Realizable $\kappa$-$\varepsilon$ and $\kappa$-$\omega$ SST model predictions, these are clearly detectable. The 2D Realizable $\kappa$-$\varepsilon$ simulations largely overpredict the magnitude of the fluctuations, while with the 3D Realizable $\kappa$-$\varepsilon$ and $\kappa$-$\omega$ SST, the magnitude deviation (compared to experiments) is notably lower (see detailed discussion in **Section 5.2.2).** The phase shift between simulation results obtained by the various numerical models and with respect to the experiments is expected in view of the variations in time (and locations) where the liquid plugs are formed in the jumper. Obviously, the mechanistic model cannot produce pressure fluctuations since it assumes a flow of a liquid plug over the entire flow domain (as described in **Section 4**), and does not consider its disintegration and the resulting cyclic acceleration-deceleration of the liquid by the air flow. Therefore, a smooth pressure decline is predicted by the mechanistic model.

The mechanistic model allows us to easily distinguish between the components of the pressure build-up, which relate to hydrostatic head, friction, and acceleration terms (see **Figure 5.7**). The first to appear is the acceleration component that shows up when the accumulated liquid is accelerated from rest. The friction component grows gradually as the liquid plug gains velocity (see initial times at stage I→II, **Figure 5.7.b1)**. Although the plug did not reach its final velocity in the tested case, the friction component reduces quite sharply when the plug moves out from the riser (i.e., the pressure drop monitored section). For the same reason, a similar steep reduction is shown for the acceleration component (see, **Figure 5.7.b3)**. Both the acceleration and friction components vanish as the liquid plug leaves the monitored section. The hydrostatic component increases while the liquid front moves into the riser and reaches a constant value when its tail leaves the lower horizontal section and thus the whole plug is in the riser (see **Figure 5.7.b2)**. The hydrostatic component starts to decrease when the plug front reaches the top elbow, and the amount of liquid in the riser reduces gradually. Finally, after the liquid tail leaves the riser's top elbow, the monitored differential pressure (that includes the contribution of the three components) decreases to zero since the model ignores the gas frictional pressure, which is practically negligible (≈0.01 atm).

In the following sections, the experimental results are compared with the predictions of the numerical simulations and the mechanistic model by focusing on the maximum pressure drop (**Section 5.2.1**) and the post-peak pressure fluctuations **(Section 5.2.2**).

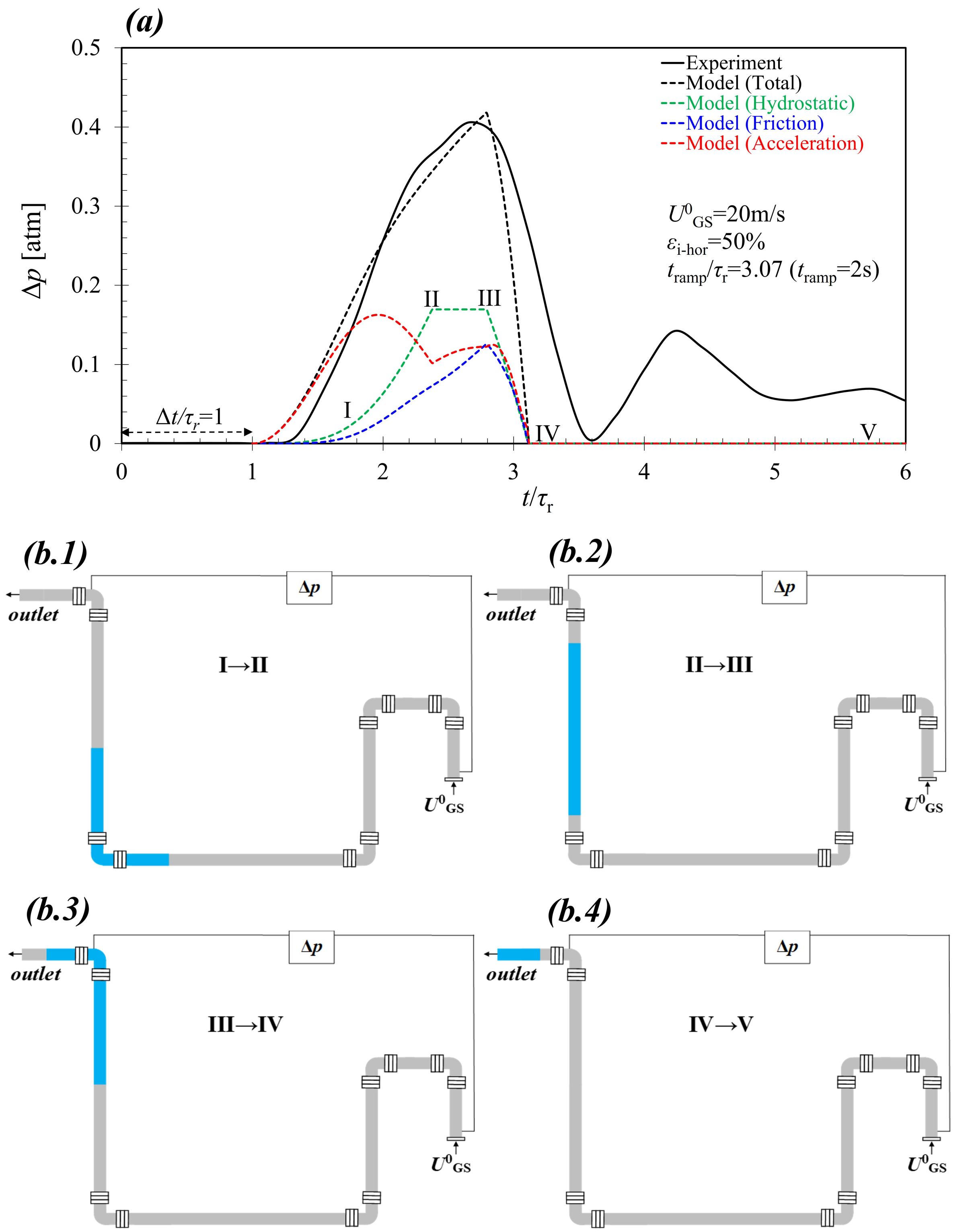

**Figure 5.7**: *(a)* Comparison between the experimental data and mechanistic model predictions (shifted by $\Delta t/\tau_r$=1 to account for plug formation delay in the experiment) for the pressure drop variation during the accumulated liquid displacement. The pressure drop components (hydrostatic, friction and acceleration) are shown, *(b)* Schematic description of the predicted liquid plug locations at the indicated time intervals.

### *5.2.1 Pressure peak*

- ***Effect of gas velocity and initial water amount***

The experiments, the mechanistic model, and the numerical simulations indicate that the first pressure drop peak, $\Delta p_{peak}$, increases with gas velocity, $U^0_{GS}$, and with the initial water amount, $\varepsilon_{i\text{-}hor}$ (see **Figure 5.8**). The numerical simulations (both 2D and 3D) reasonably predict the experimental pressure drop peak for the different gas velocities and initial water amounts. The initial pressure rise depends on the gas-liquid two-phase flow characteristics (for given $\varepsilon_{i\text{-}hor}$ and $U^0_{GS}$). In most cases (particularly for high $U^0_{GS}$ or/and high $\varepsilon_{i\text{-}hor}$), a plug of liquid is created at the initial stages of the gas flow (see **Figure 5.3**). The plug is accelerated, and most of the liquid volume leaves the jumper. During this process, the pressure drop rises to its peak and then declines. The plug creation in the simulations is crucial for a correct prediction of the pressure drop. Some numerical models (i.e., 2D &3D with standard $\kappa$-$\varepsilon$, see **Figure 5.3.c&f1**) showed a prolonged stratified flow pattern in the low horizontal section, and liquid blockage appeared only close to the riser's bottom elbow at later flowtimes. Therefore, those numerical models usually underpredict the pressure drop peak compared to other numerical setups (i.e., 2D $\kappa$-$\omega$ SST and 2D&3D Realizable $\kappa$-$\varepsilon$).

The mechanistic model predictions show good agreement with the experiments for high gas velocities and large initial liquid amounts. For those cases, the model's assumption that the whole amount of liquid forms a plug already at the initial stages of the liquid displacement is appropriate. However, the experimental observations reveal that with low initial liquid amounts (and with low gas velocities), a plug that contains only a small fraction of the liquid may be created only near the riser elbow (or in the riser) section. Then, the initial plug rapidly disintegrates in the riser by the gas flow. In contrast, this plug disintegration is not considered by the mechanistic model. For this reason, the mechanistic model predictions overpredict the experimental results at low gas velocities and small liquid amounts. Comparing the mechanistic model predictions with the experimental data suggests that the model is valid for $\varepsilon_{i\text{-}hor}>50\%$ and $U^0_{GS}>15$m/s.

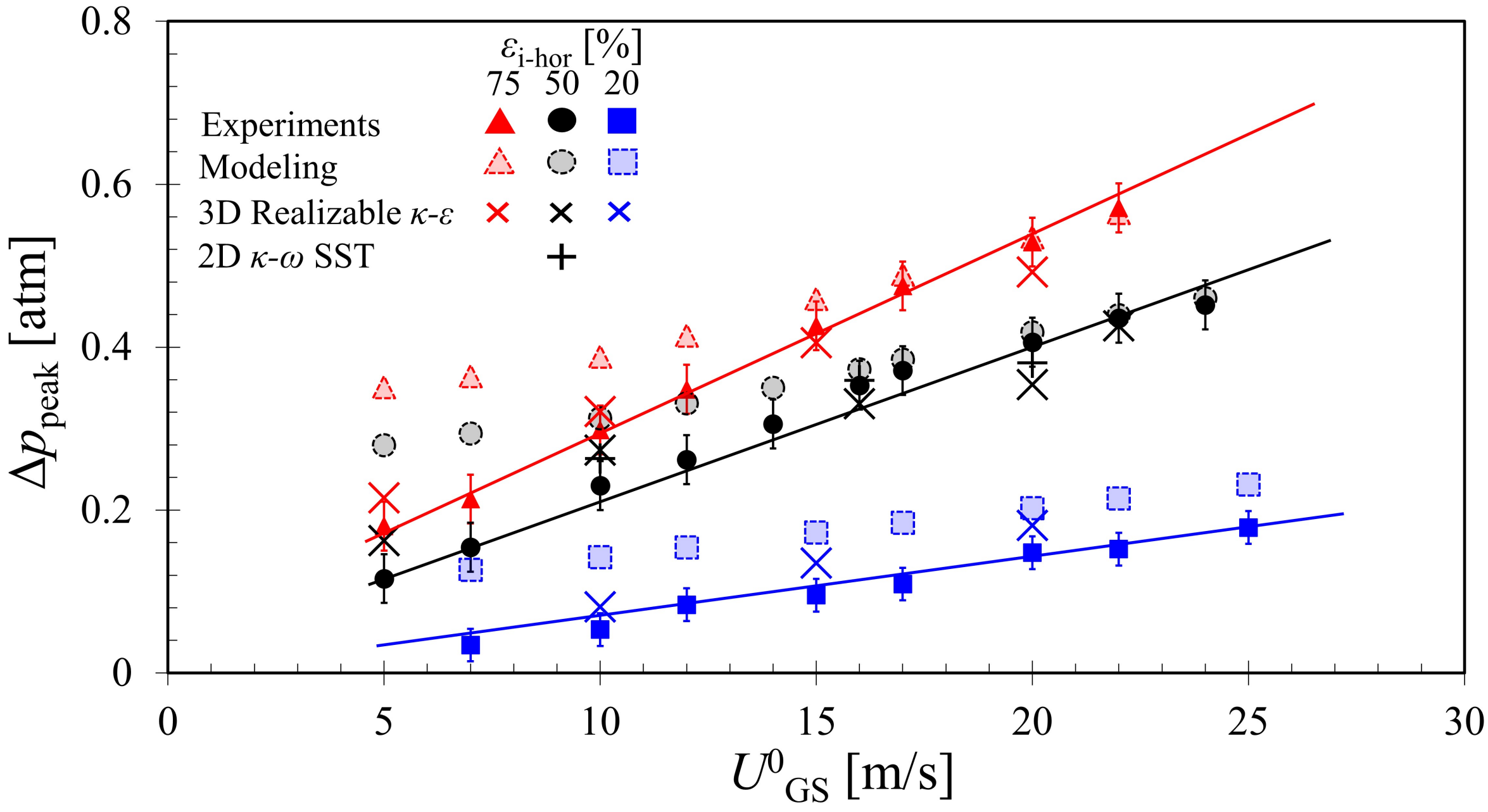

**Figure 5.8**: Comparison between the experiments, mechanistic modeling and numerical simulations regarding the gas pressure drop peak, $\Delta p_{\mathrm{peak}}$ vs. the gas velocity, $U^0_{\mathrm{GS}}$, for various initial water amounts, $\varepsilon_{\mathrm{i\text{-}hor}}$. Ramp-up period $t_{\mathrm{ramp}}$=2 s ($t_{\mathrm{ramp}}/\tau_{\mathrm{r}}$=0.76÷3.84, for $U^0_{\mathrm{GS}}$=5÷25m/s).

- ***Effect of the ramp-up period***

Prolonging the ramp-up period may moderate the liquid acceleration rate, suggesting that the pressure peak can be controlled by the ramp-up period. **Figure 5.9** shows the effect of the ramp-up period on the pressure drop peak. Indeed, doubling the ramp-up period leads to a decrease of the initial pressure drop peak by up to ~50% (experiments and numerical simulations). The same effect was observed for low gas flow rate ($U^0_{\mathrm{GS}}$=10 m/s) and high gas flow rate ($U^0_{\mathrm{GS}}$=20 m/s). Importantly, the ramp-up period does not affect the critical gas velocity (≈20 m/s). It means that extending the ramp-up period assures safer water removal.

**Figure 5.9** presents a comparison between results obtained in the experiments, by the mechanistic model, and the 3D numerical simulations for various ramp-up periods and initial water volumes. The numerical simulations show good agreement with the experimental data for all tested cases. However, the mechanistic model, which assumes liquid plug flow, consistently over-predicts both the experimental and numerical results. As discussed with reference to **Figure 5.8**, the deviations are higher for low initial liquid amounts and low gas superficial velocities as the model does not consider the liquid phase disintegration during its displacement by the gas flow. This deficiency of the model obviously becomes more significant as the residence time of the liquid in the jumper (affected by a longer ramp-up) increases. Therefore, the presented deviations of the mechanistic model predictions from the experimental (and numerical) data are somewhat expected.

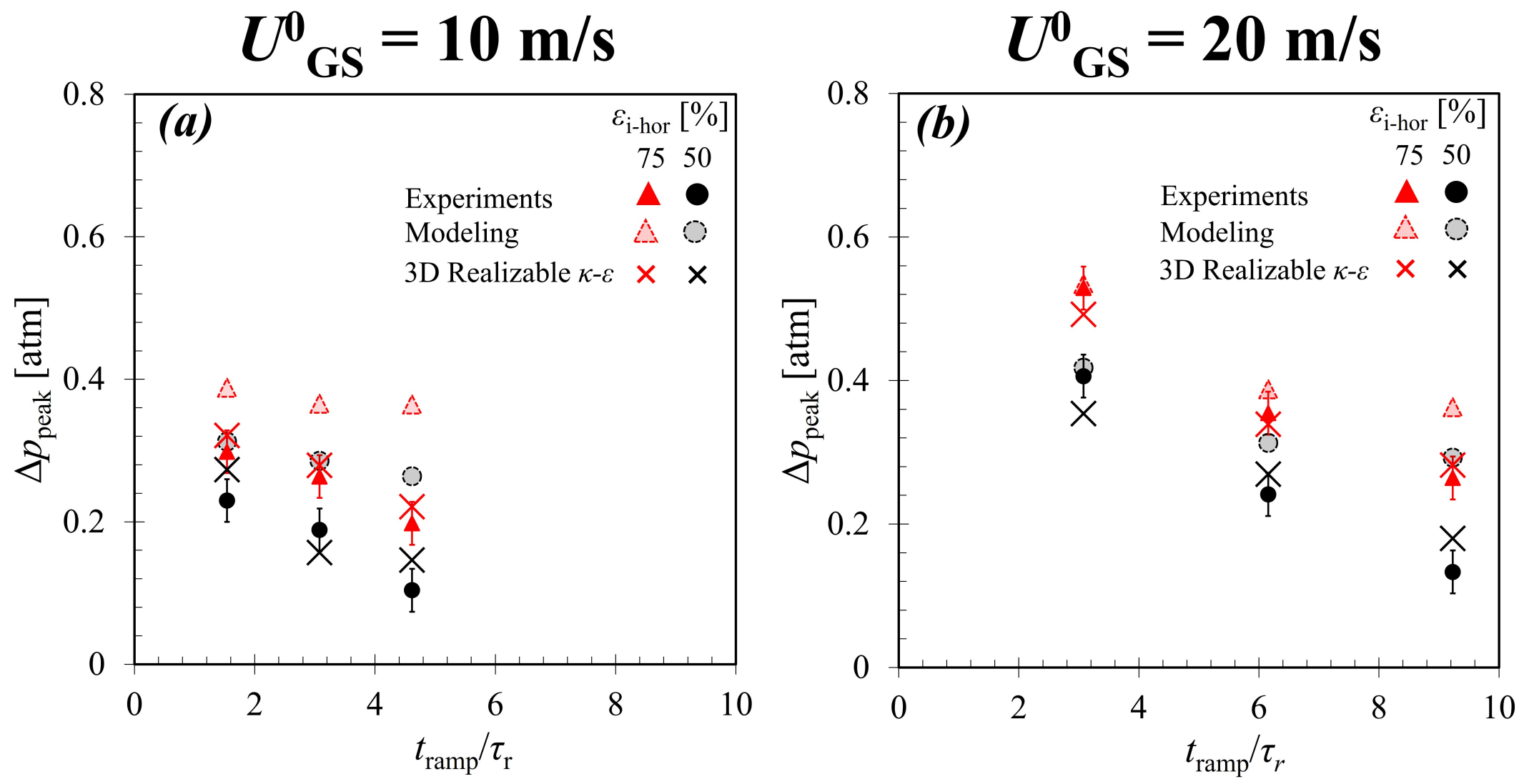

**Figure 5.9**: Comparison between the experiments, mechanistic modeling and numerical simulations regarding the gas pressure drop peak, $\Delta p_{\text{peak}}$, vs. the normalized (by the gas residence time) ramp-up period, $t_{\text{ramp}}/\tau_{\text{r}}$, for various initial water amounts, $\varepsilon_{\text{i-hor}}$. $U^0_{\text{GS}}$= *(a)* 10m/s and *(b)* 20 m/s.

### *5.2.2 Pressure fluctuations*

The experiments and the numerical simulations (see **Figure 5.6**) reveal that periodic pressure variations at later flowtimes are caused by cyclic creation and breakage of liquid plugs near the lower riser elbow or farther in the riser. Although such behavior was observed for sub- and super-critical gas velocities, the associated flow phenomena might be different. Note, however, that in both cases, some of the liquid will be removed from the jumper during the initial pressure rise period (i.e., initial plug creation).

At subcritical gas velocities (i.e., $U^0_{\text{GS}}<U^0_{\text{GS|Crit}}$), the gas velocity is not sufficiently high to enable steady annular concurrent up-flow, where backflow in liquid film on the riser walls is avoided. As a result of the backflow, some of the liquid accumulates in the lower riser elbow, whereby a liquid bridge that results in temporary airflow blockage is formed. In addition, the liquid is not completely displaced from the horizontal section at the initial stages, thus it arrives only later to the riser elbow region and may also cause a blockage to gas flow. As a result, the gas pressure rises to displace the liquid upwards (and partially remove it from the jumper) and declines upon aeration/disintegration of the liquid bridge. The process repeats as long as the residual liquid amount is large enough to sustain a cyclic blockage formation. During the initial cycles, fractions of the liquid are removed repeatedly from the system, so both the amplitude and the average of the pressure fluctuations decrease with time. Eventually, the pressure reaches a steady cyclic behavior. In the experiments, at very long flowtimes, the evaporation of the liquid into the air becomes significant, and the pressure fluctuations may further decay with the associated reduction of the liquid amount in the riser and in its elbow region.

At super-critical gas velocity (i.e., $U^0_{GS}>U^0_{GS|Crit}$), the liquid backflow in the riser is prevented. However, pressure fluctuations are still observed similarly to cases with $U^0_{GS}<U^0_{GS|Crit}$. Those are caused by the delayed arrival of liquid that was not flushed from the horizontal section already at the initial stages, to the riser. There, the liquid repeatedly blocks the gas flow. However, with $U^0_{GS}>U^0_{GS|Crit}$ the pressure fluctuations decrease rapidly as the super-critical gas flow removes the liquid. Eventually, the pressure reaches a steady (relatively low) constant value (corresponding to the frictional pressure drop along the jumper).

The pressure fluctuation frequencies resemble the liquid bridge creation frequencies, and those are similar (~1 Hz) for all tested cases in the experiments and the numerical simulations (2D&3D). Also, this frequency is practically independent of the initial liquid amount and the gas velocity (in the tested ranges). However, the measured fluctuation amplitudes are usually overpredicted by the numerical simulations (3D&2D Realizable $\kappa$-$\varepsilon$ and 2D $\kappa$-$\omega$ SST). As mentioned, the experimental observations reveal that the liquid tends to disintegrate during its displacement by the gas. The liquid breakage is most notable in the riser section. Similar patterns were also detected in the numerical simulations. However, the VOF algorithm is known to be less effective for simulating dispersed flows (i.e., non-segregated flows without distinct interface between the phases, e.g., Gregor et al., 2000; Hänsch et al., 2013; Wardle and Weller, 2013b; Yurishchev et al., 2022). Therefore, the flow patterns in the experiments and the numerical simulations might appear very similar, yet the fine two-phase dynamics are different. The results suggest that the liquid bridges created in the numerical simulations are larger than those created in the experiments, resulting in larger liquid lumps that need to be lifted in each cycle and higher pressure fluctuations amplitude.

### 5.3 Forces

The jumper elbows are exposed to dynamic loads during the passage of liquid through their elbows. As elaborated in the previous sections, a liquid plug may be formed during transient conditions. The liquid plug is accelerated by the gas while flowing in the riser and then hits its upper elbow. Forces acting on the riser's upper elbow are translated to moments acting on the lower elbows. Therefore, in the following, the forces exerted on the riser's upper elbow are examined. Also, this region bears the highest loads during the transient displacement of the liquid, as the liquid arrives at a relatively high velocity.

The force obtained from the mechanistic model is based on the momentum balance on the liquid in the elbow (as elaborated in **Section 4**) and consists of two components due to the pressure and the liquid inertia. The 3D numerical simulation, however, enables calculating directly the force acting on the elbow (of its real geometry) through the integral of the gauge pressure acting on its surface (the outer pressure is atmospheric). The force can also be calculated via the mechanistic model approach, based on a momentum balance on the fluid passing through the elbow, taking the pressure and inertia components derived by integrating the local values obtained in the 3D simulations at the outlet cross-section. A comparison of the forces obtained from the simulation results via these two approaches with the predictions of the mechanistic model enables testing the impact of the model assumptions. **Figure 5.10** shows that the force calculated by integrating the gauge pressure on the elbow surface (in blue) is similar to the force obtained by considering the pressure and the liquid inertia at the elbow outlet (in red). Moreover, the neglect of the wall friction appears to be justified. As shown in **Figure 5.10**, moving away the elbow outlet cross-section ($4D$ downstream of the actual elbow outlet, in green), practically does not affect the resulting force experienced by the elbow. It is worth noting that for transient two-phase flow, the momentum in the elbow is a function of time and space since both the velocity and the density are time and space dependent. Therefore, the time derivative of the $x$-momentum, $\partial/\partial t(\int \rho U_x \mathrm{d}V)$, should, in principle, be included in the momentum balance on the fluid in the elbow. This results in an additional contribution to the horizontal force. Comparing the force so-obtained (in black) to that obtained when the $\partial/\partial t(\int \rho U_x \mathrm{d}V)$ term is neglected (in red, **Figure 5.10**) reveals that the contribution of this term is rather insignificant. The peak value of the force predicted by the mechanistic model shows a reasonable agreement with that obtained by the numerical simulations.

The mechanistic model assumptions lead to $F_x = F_y$ (see **Section 4**). This result was verified via numerical simulations. **Figure 5.11.a** shows the force components (in $x$, $y$, and $z$ directions) acting on the riser's upper elbow for a specific case. It is clear that $F_x$ (solid red) $\approx F_y$ (dashed blue). The forces in $z$ direction ($F_z$, solid black) may arise due to asymmetry of the pressure distribution on the elbow surface. Nevertheless, those forces are of much lower magnitude (compared to the $x$, $y$ components) and do not show a particular pattern.

The accumulated liquid passes through three elbows (points 5, 6, and 7 in **Figure 4.1**) on its way to the jumper outlet. Each of the elbows experiences a force peak as the liquid reaches it. In view of the results in **Figure 5.11.b**, which indicate that elbow 6 bears the most significant forces, the decision to measure the horizontal force acting on this elbow appears justified. Indeed, the horizontal forces ($x$-direction) on elbows 5 and 7 (black and blue lines) are lower than on elbow 6 (red line). As expected, the time delay between the force peaks in elbows 6 and 7 is rather small, but the peak force on elbow 7 is about half of the peak force on elbow 6. This is a result of the continued aeration/disintegration of the liquid lump during its flow in the horizontal section connecting the two elbows, and thereby, the reduction of the fluid momentum upon reaching elbow 7.

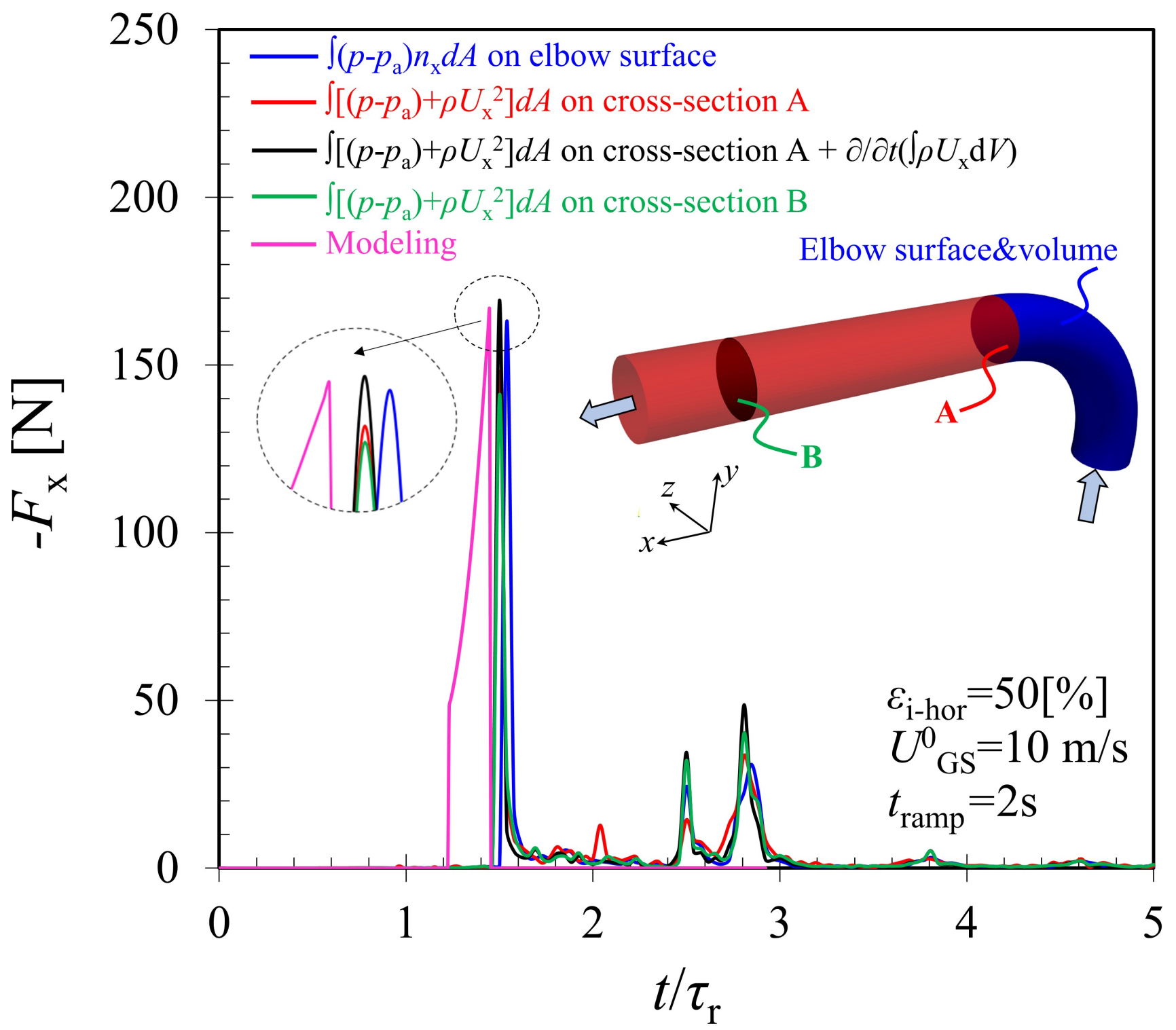

**Figure 5.10:** The time variation of the horizontal force acting on the riser's upper elbow during the purge-out of accumulated water ($\varepsilon_{i\text{-}hor}$=0.5 and ramp-up period $t_{ramp}$=2 s, corresponding to $t/\tau_r \approx$ 1.54 for $U^0_{GS}$=10 m/s). Comparison between the mechanistic model prediction and numerical simulations results.

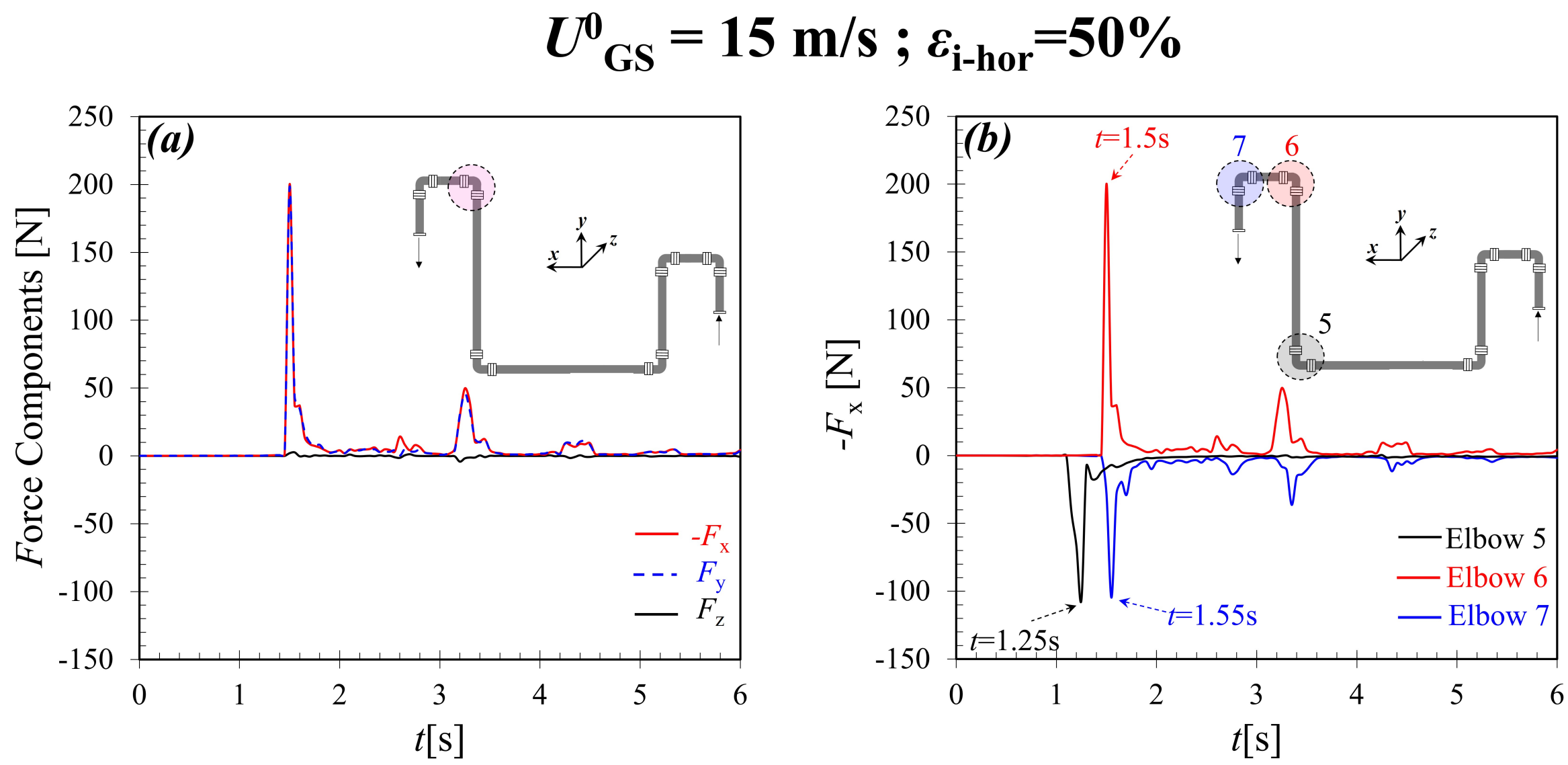

**Figure 5.11**: The time variation of ($a$) the force components acting on the riser's upper elbow 6 and ($b$) the force in the x-direction acting on the riser's elbows 5, 6 and on elbow 7 during the displacement of the accumulated water, for $U^0_{GS}$=15 m/s, $\varepsilon_{i\text{-}hor}$=0.5 and ramp-up period $t_{ramp}$=2s.

**Figure 5.12** shows examples of the horizontal force component acting on the riser's upper elbow, $F_x$, vs. the time for different gas velocities, $U^0_{GS}$ ($t_{ramp}$=2 s). A short-duration excessive force was measured in the experiments, with a peak value and timing similar to that calculated by the numerical simulations and predicted by the mechanistic model (except at low subcritical $U^0_{GS}$ = 5 m/s, where the force peak in the simulations is somewhat delayed (by ~ 0.5s) compared to the experimental results). However, the force variation in the later times is different. In the numerical simulation results, the initial force peak is followed by a series of weaker peaks caused by the acceleration of some additional residual liquid chunks. Those later force peaks do not exhibit a particular pattern (or frequency). Yet, experiments showed clear cyclic force variations with a frequency of ~5 Hz after the initial (and significant) force peak. The transient two-phase flow in the elbow may not be the lone factor of the force fluctuations. Apparently, this behavior is related to the dynamic response of the experimental elastic jumper (made of Perspex) and its supporting aluminum structure, which exhibit large vibrations following the impact of the water on the elbow that endanger the structure already when operating close to the critical $U^0_{GS|Crit}$. These structure vibrations mask the dynamic variation of the force on the elbow due to two-phase flow dynamics (such as the arrival of water chunks into the elbow), which are observed in the simulations results.

**Figure 5.13** demonstrates the horizontal force peak as a function of gas velocity $U^0_{GS}$ (with $t_{ramp}$=2s) for various initial liquid amounts, $\varepsilon_{i\text{-}hor}$. The results show a reasonable agreement between the experimental data, numerical simulation results, and the mechanistic model predictions for most cases when the force increases with gas velocity. As expected, the force exerted on the elbow increases with the gas velocity. As previously discussed, the mechanistic model tends to overestimate the maximal force for low gas velocities. Apparently, for such cases, the single plug assumption might be less realistic.

**Figure 5.14** describes the horizontal force peaks vs. the normalized ramp-up period, $t_{ramp}/\tau_r$, for various initial liquid amounts. For longer $t_{ramp}/\tau_r$ (i.e., lower ramp-up rate), the maximal force is smaller. This suggests that the maximal force can be moderated by reducing the ramp-up rate, thereby avoiding risk to the jumper structure. Here, the overprediction of the force obtained by the mechanistic model increases with $t_{ramp}/\tau_r$. In a prolonged ramp-up period, a (single) liquid plug assumed in the model may not comply with the actual liquid flow pattern in the jumper.

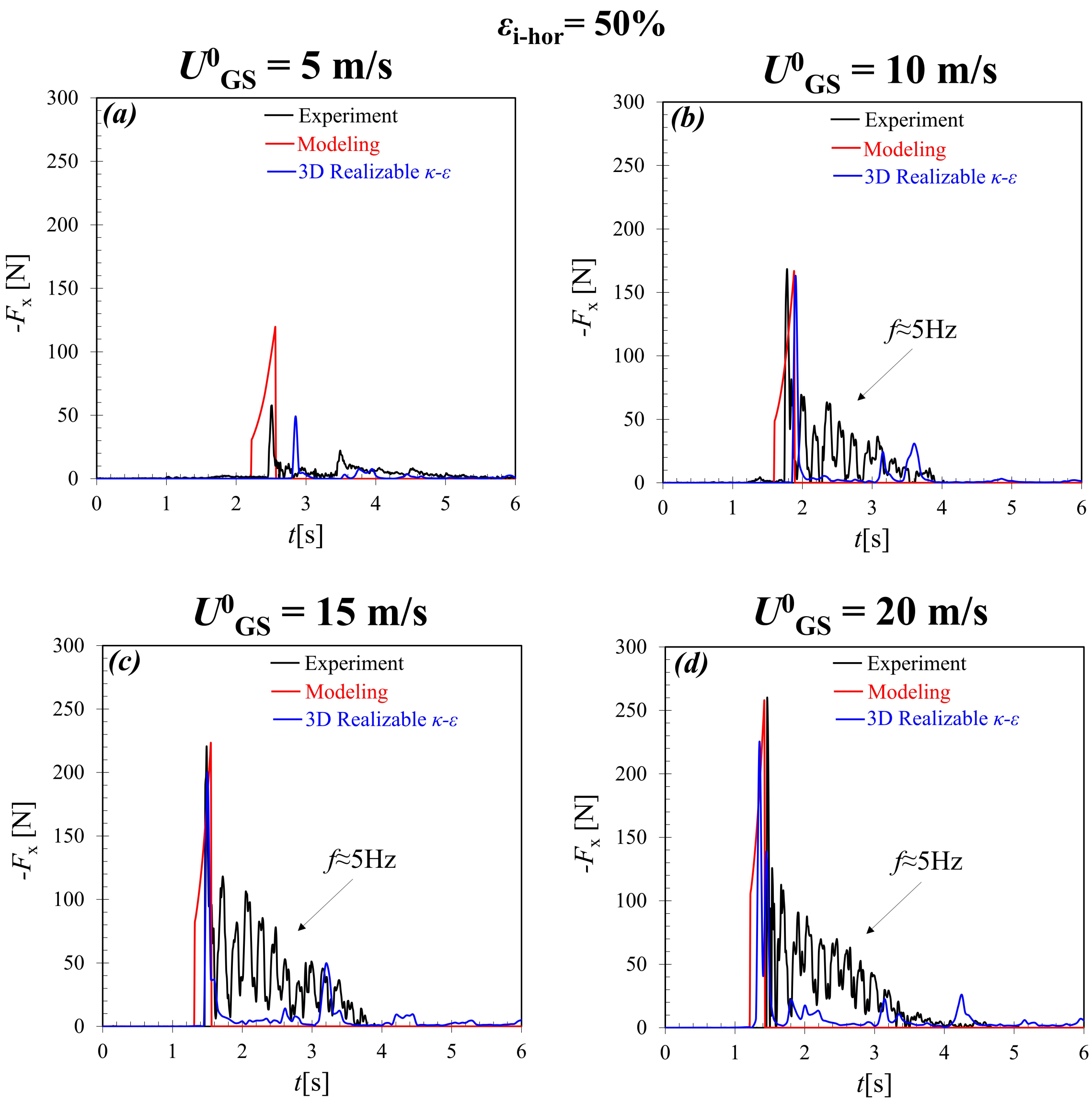

**Figure 5.12**: Time variation of the horizontal force on the riser upper elbow during the purge-out of accumulated water ($\varepsilon_{i\text{-}hor}$=0.5 and ramp-up period $t_{ramp}$=2 s). Comparison of results obtained by experiments with numerical simulations (3D Realizable $\kappa$-$\varepsilon$) and the mechanistic model predictions (the experimental error is provided in **Table 2.1**).

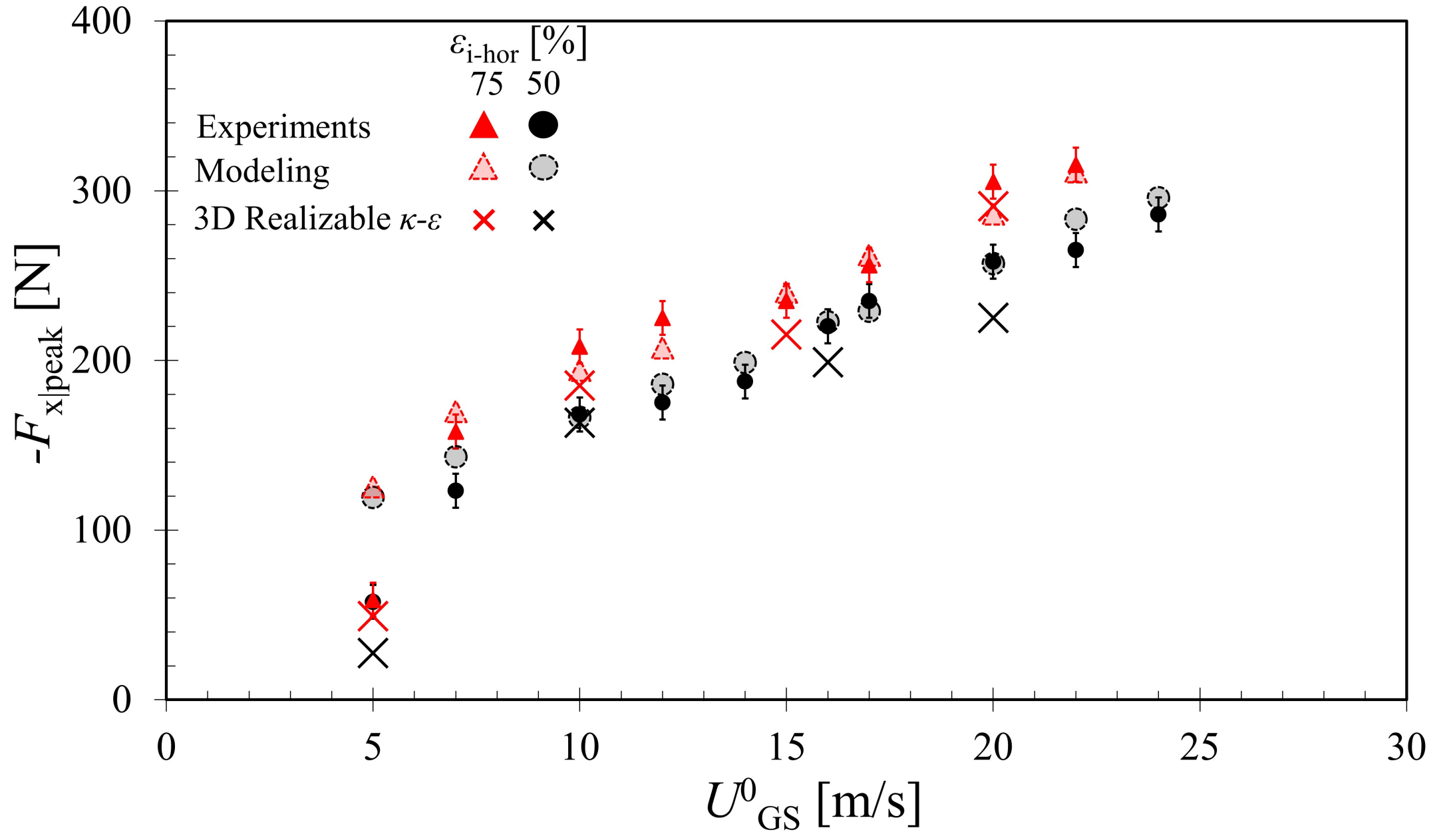

**Figure 5.13***:* Maximal horizontal force on the riser upper elbow, $\Delta F_{x|peak}$ vs. the gas velocity, $U^0_{GS}$, for various initial water amounts, $\varepsilon_{i\text{-}hor}$: Comparison between the experiments, mechanistic modeling, and numerical (3D Realizable $\kappa$-$\varepsilon$) simulations. Ramp-up period $t_{ramp}$=2 s ($t_{ramp}/\tau_r$=0.76÷3.84, for $U^0_{GS}$=5÷25m/s).

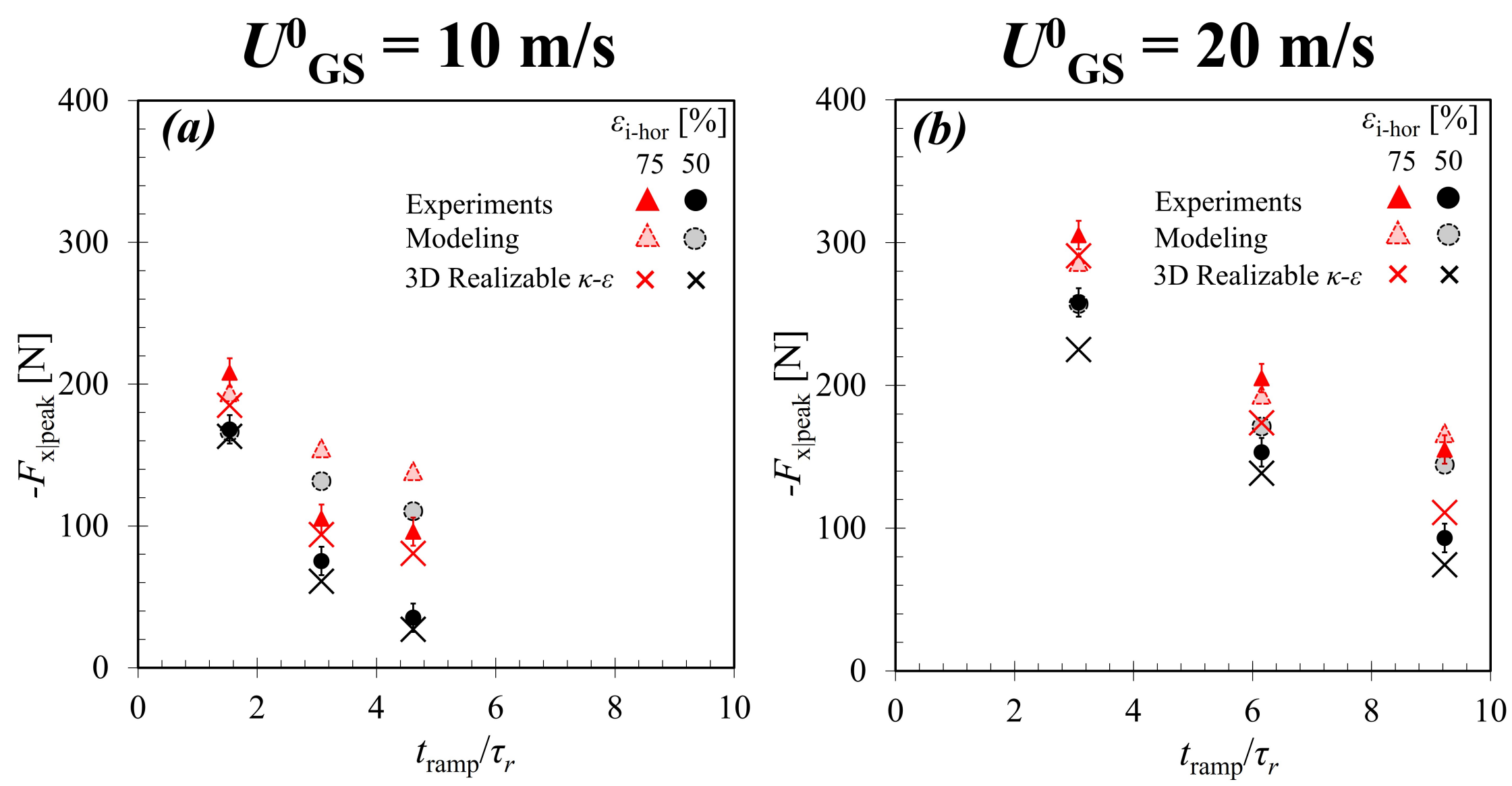

**Figure 5.14***:* Maximal horizontal force acting on the riser upper elbow, $\Delta F_{x|peak}$ vs. the normalized ramp-up period, $t_{ramp}/\tau_r$, for various initial water amounts, $\varepsilon_{i\text{-}hor}$. Comparison between the experiments, mechanistic modeling and numerical (3D Realizable $\kappa$-$\varepsilon$) simulations. *(a)* $U^0_{GS}$=10 m/s and *(b)* $U^0_{GS}$=20 m/s.

## 6 Conclusions

The study aims to investigate transient two-phase air-water flow in an M-shaped jumper, a typical tie-in structure of subsea gas production pipelines. Experiments in a scaled-down jumper and numerical simulations were combined to determine the critical gas mass-flow rate for the complete removal of liquid accumulated in the jumper. Time variation of pressure drop in the jumper and forces acting on the riser upper elbow were also examined, and a simplified mechanistic model for their prediction was established.

Experiments indicated that the critical gas flow rate (and the associated superficial velocity at standard conditions, $U^0_{GS|Crit}$) is the minimal value required for complete dry-out of the lower riser elbow, which is enabled when liquid backflow in the jumper's riser is prevented. A value of $U^0_{GS|Crit}$ ~ 20 m/s was documented in the experimental set-up, which was found to be independent of the accumulated liquid volume (initially introduced into the system). Similar results were obtained by the numerical simulations when accounting for the gas compressibility and employing a proper RANS turbulence model (3D simulations with Realizable $\kappa$-$\varepsilon$, or 2D simulations with the $\kappa$-$\omega$ SST).

The displacement of the accumulated liquid is associated with an increase in the pressure drop in the jumper, which is characterized by a pressure peak in the gas inlet at the initial stage of the water displacement. Also, the accelerated liquid impacts the jumper elbows, in particular, the riser's upper elbow, and produces a short-term and significant force. As expected, the maximal pressure drop, $\Delta p_{peak}$, and the impact force on the riser elbows increase with the gas velocity, $U^0_{GS}$, and with the accumulated liquid amount in the system represented by $\varepsilon_{i\text{-}hor}$. Both can be moderated by reducing the gas-flow ramp-up rate (i.e., increasing the ramp-up time, $t_{ramp}$), thereby reducing the risk to the jumper structure. In all the tested cases, a favorable agreement was found between the experiments and the 3D numerical simulations for the pressure and force transients and their peak values. The numerical simulations showed that the riser's upper elbow experiences the most significant load compared to the other elbows. In all elbows, the magnitude of the vertical force component, $F_y$, is practically equal to $F_x$, while the magnitude of the $F_z$ component is much smaller ($F_z << F_x$, $F_y$).

Pressure fluctuations were documented both in the experiments and the numerical simulations. The detected pressure fluctuations decay over several cycles as the water is purged from the system. Significant low-frequency (~1 Hz) pressure fluctuations were observed both in the experiments and in the simulations. This frequency was found to be independent of the initial volume of accumulated water.

The new mechanistic model assumes that the accumulated liquid is displaced as a single plug. The model is shown to be a useful simple tool to predict the pressure transient and forces on the elbow for close to critical (and super-critical) $U^0_{GS}$ and large amounts of accumulated liquid ($\varepsilon_{i\text{-}hor} \sim 1$). However, the model tends to overestimate the pressure and force peak values for relatively low $U^0_{GS}$, small amounts of accumulated liquid, and long ramp-up time. The model over-prediction is attributed to the aeration and/or disintegration of the liquid slug during its displacement. Under such conditions, those phenomena become dominant, and the model predictions should be considered an upper bound. Nevertheless, the advantage of the model (in addition to its simplicity) is its ability to identify the significance of the various components that contribute to the pressure variation (hydrostatics, friction, and acceleration), thereby enabling the interpretation of the experimental and simulation results.

In the course of the study, the applicability of the standard and Realizable $\kappa$-$\varepsilon$ (and $\kappa$-$\omega$ SST) turbulence models was tested and verified against the experimental data. The discussed (apparently inaccurate) results obtained by applying the standard $\kappa$-$\varepsilon$ turbulence model are somewhat expected. The model is essentially a *high*-Re formulation, which is typically suitable for turbulent developed flows. Therefore, it may not be adequate for the simulation of the transient flow considered in the current study, where initially, the fluids are stationary, and then the air flow rate gradually increases to its terminal value. In contrast, the $\kappa$-$\omega$ SST and Realizable $\kappa$-$\varepsilon$ models are considered as *low*-Re turbulence models and consequently may be a better choice. Indeed, the simulated cases on the 3D domain with Realizable $\kappa$-$\varepsilon$ and the 2D domain with $\kappa$-$\omega$ SST produced reliable results in terms of the gas pressure response and the critical gas velocity for water removal. We believe that 3D simulations with $\kappa$-$\omega$ SST would produce the most accurate predictions. Unfortunately, such calculations were not carried out due to computational capability limitations. Obviously, simulations in a 2D domain do not enable the correct representation of the phases' distribution and dynamics in the vertical riser, and it was impossible to reproduce the experimentally observed churn and annular flow patterns with the 2D simulations. Hence, it is recommended to perform future parametric studies (i.e., sub-sea site gas pressures and pipe diameters) with the 3D Realizable $\kappa$-$\varepsilon$ model, for which the numerical tool results were found to reasonably agree with the experimental data.

## Acknowledgment

This study was supported through generous funds provided by the Israel Ministry of Energy under contract number 220-17-005.

## Appendix A: grid and time convergence tests

Grid and time convergence tests were conducted to assess the numerical accuracy and stability of the computational models employed. The grid convergence tests involved refining the computational mesh while keeping all other parameters constant and observing the changes in the solution. Multiple grid resolutions were investigated, ranging from coarse to fine (3 grids for the 2D model and 4 grids for the 3D model, see **Table A.1**). Initially, the grid convergence tests were carried out with a maximum Courant number of 0.5. To further evaluate the impact of temporal discretization, the selected grid was subjected to sensitivity analysis by decreasing the maximum Courant number to 0.25. The described methodology was repeated for 3 turbulence models (standard and Realizable $\kappa$-$\varepsilon$, and $\kappa$-$\omega$ SST). This is to demonstrate the difference between results affected by the selected turbulence model, rather than by the spatial or temporal discretization method.

The convergence tests (grid and time) focused on the main parameters that were investigated in the present study (1) the critical gas velocity needed to purge the accumulated liquid (see **Figure A.1**) (2) the pressure drop on the Jumper during the liquid displacement (see **Figure A.2** and **Figure A.3**), and (3) the horizontal force applied on the riser's upper elbow during the liquid passage (see **Figure A.4**). The tests revealed that further grid refinement (relatively to the selected grid) and reducing the maximal Courant number to 0.25 (from 0.5), did not significantly affect the results.

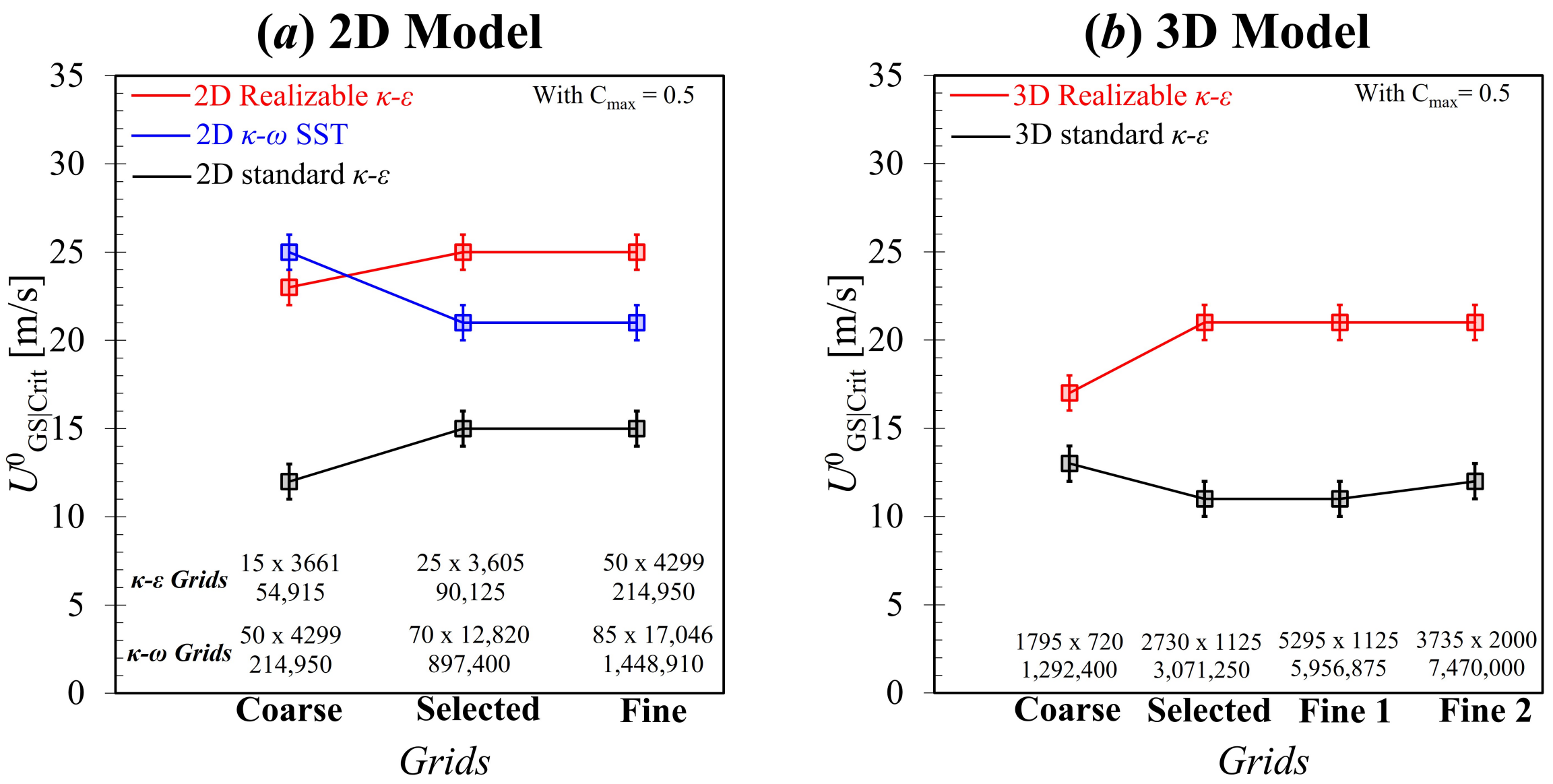

**Figure A.1:** Grid convergence tests regarding the critical gas velocity needed to purge the accumulated liquid from the Jumper for (*a*) 2D Model and (*b*) 3D Model, while applying different turbulence models (standard and Realizable $\kappa$-$\varepsilon$, and $\kappa$-$\omega$ SST).

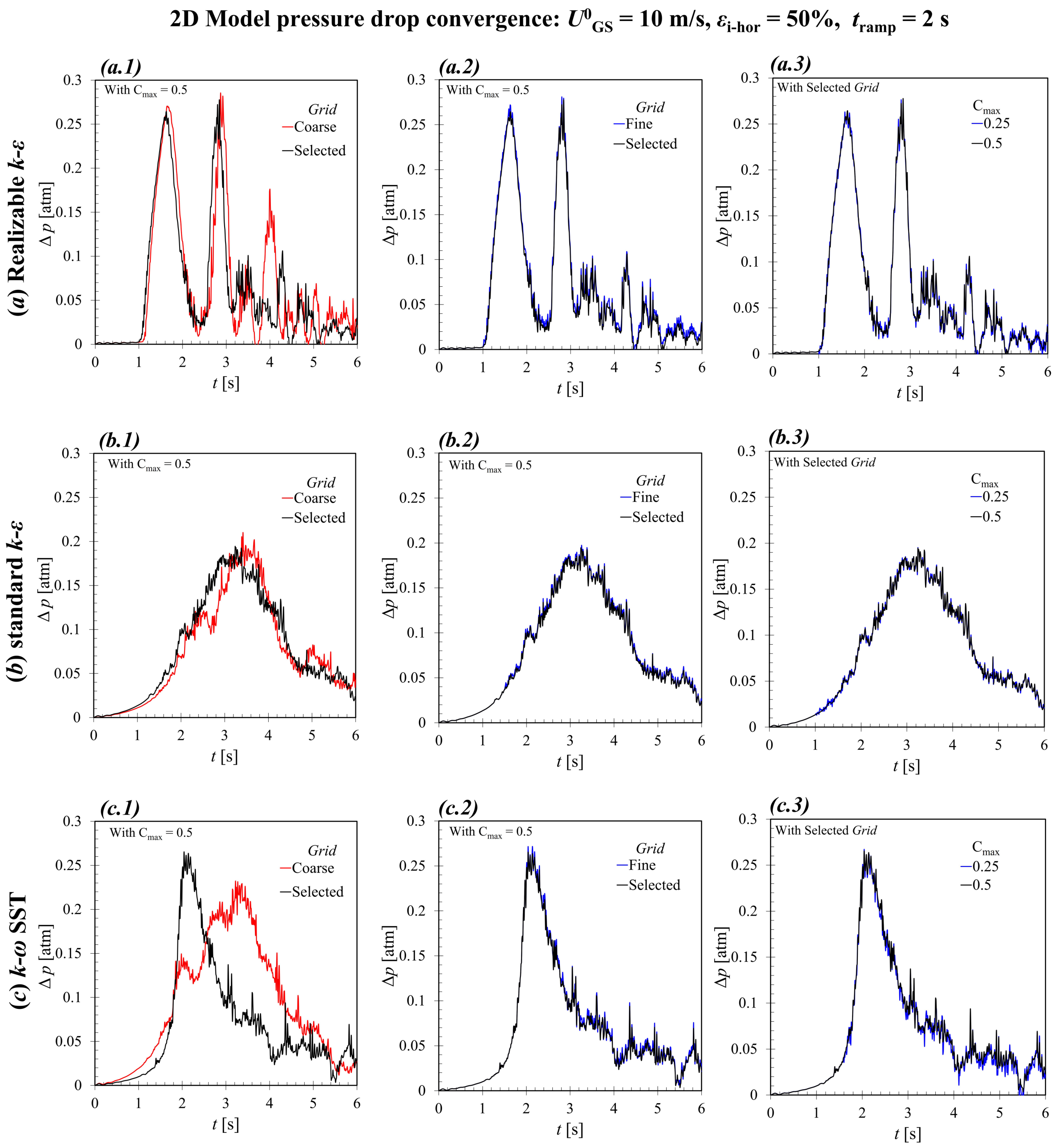

**Figure A.2:** Grid and time convergence tests regarding the pressure drop for 2D model for a specific case while applying (*a*) Realizable κ-ε, (*b*) standard κ-ε or (*c*) κ-ω SST turbulence models.

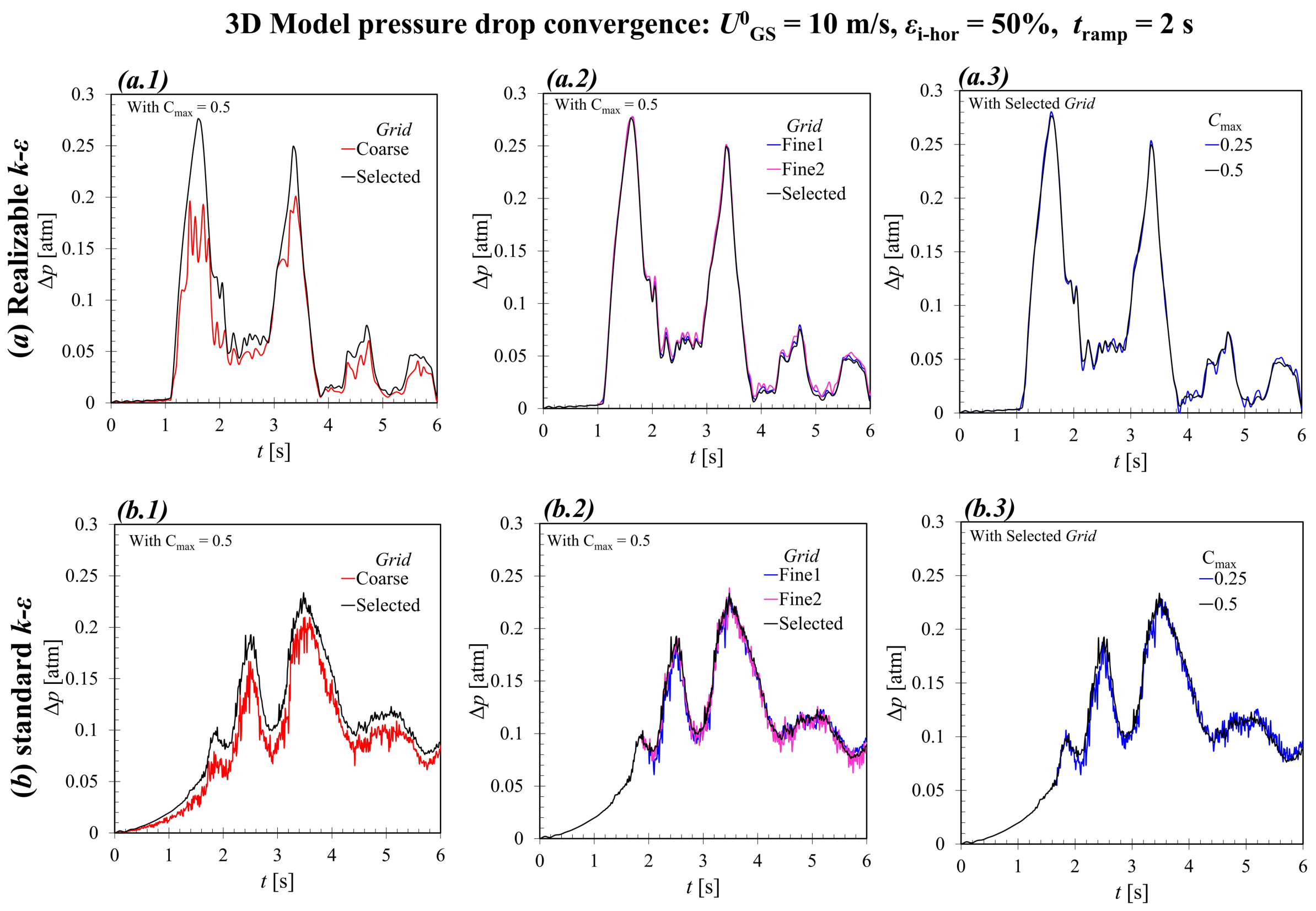

**Figure A.3:** Grid and time convergence tests regarding the pressure drop for 3D model for a specific case while applying (*a*) Realizable *κ*-*ε* or (*b*) standard *κ*-*ε* turbulence models.

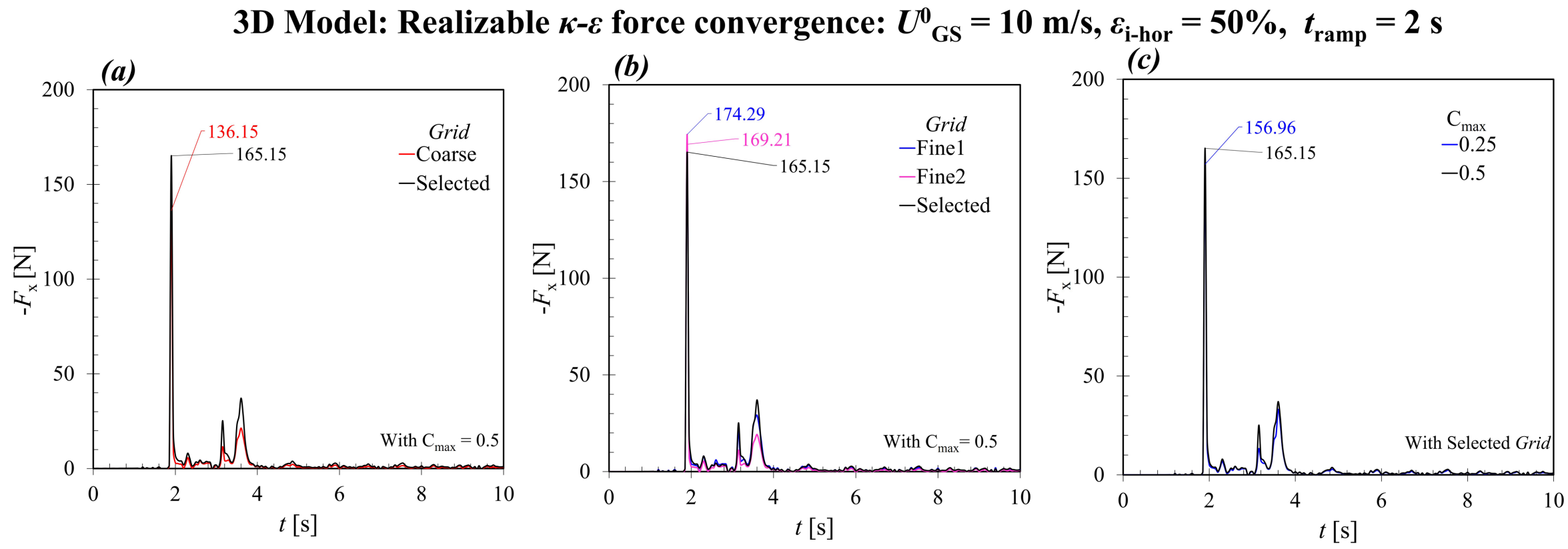

**Figure A.4:** Grid and time convergence tests regarding the horizontal force acting on the riser's upper elbow convergence.

**Table A.1:** Lateral, axial, and total element count for the grids used in grid convergence tests.

| ***Grids*** | ***Axial* x *Lateral elements (Number of elements)*** | | |
|---|---|---|---|
| | 3D *κ*-*ε* | 2D *κ*-*ε* | 2D *κ*-*ω* |
| Fine 1 | 5,295 x 1,125 (5,956,875) | - | - |
| Fine 2 | 3,735 x 2,000 (7,470,000) | 50 x 4,299 (214,950) | 85 x 17,046 (1,448,910) |
| Coarse | 1,795 x 720 (1,292,400) | 15 x 3,661 (54,915) | 50 x 4,299 (214,950) |
| Selected | 2,730 x 1,125 (3,071,250) | 25 x 3,605 (90,125) | 70 x 12,820 (897,400) |

## Appendix B: discretization error estimation

The Grid Convergence Index (GCI) method was applied to estimate the discretization error in the present study. The method serves as a measure of the percentage by which the computed value deviates from the value of the asymptotic numerical solution, and thereby offers insights into how the solution might change with further grid refinement. The GCI method is based on Richardson, 1911; Richardson and Gaunt, 1927 .In the following, the recommended procedure for estimating the discretization error presented in Celik et al., 2008 is applied.

A representative cell size, $h$ for the grid used in the grid convergence tests is defined (for more details see **Appendix A.**

$$h = \left[\frac{1}{N}\sum_{\mathrm{i}=1}^{N} \Delta V_{\mathrm{i}}\right]^{\frac{1}{3}} \qquad \text{[B.1]}$$

where $\Delta V_{\mathrm{i}}$ is the volume the i$^{\mathrm{th}}$ cell, and $N$ is the total number of cells used for the computations.

The apparent order of the method accuracy, $p$ is given by:

$$p = \frac{1}{\ln r_{21}}\left|\ln\left|\frac{\varepsilon_{32}}{\varepsilon_{21}}\right| + q(p)\right|$$

$$q(p) = \ln\left(\frac{r_{21}^{p} - s}{r_{32}^{p} - s}\right) \qquad \text{[B.2]}$$

$$s = 1 \cdot \mathrm{sign}\left(\frac{\varepsilon_{32}}{\varepsilon_{21}}\right)$$

where $r_{21}=h_2/h_1$ and $r_{32}=h_3/h_2$, with $h_1<h_2<h_3$, and $\varepsilon_{21} = \varphi_2 - \varphi_1$, $\varepsilon_{32} = \varphi_3 - \varphi_2$, with $\varphi_{\mathrm{k}}$ denoting the solution on the k$^{\mathrm{th}}$ grid. The extrapolated value resulting from further grid refinement, $\varphi_{\mathrm{ext}}^{21}$ is then:

$$\varphi_{\mathrm{ext}}^{21} = \frac{\left(r_{21}^{p}\varphi_1 - \varphi_2\right)}{\left(r_{21}^{p} - 1\right)} \qquad \text{[B.3]}$$

The corresponding relative errors, $e_{\mathrm{a}}^{21}$ and $e_{\mathrm{ext}}^{21}$ are:

$$e_{\mathrm{a}}^{21} = \left|\frac{\varphi_1 - \varphi_2}{\varphi_1}\right|$$

$$e_{\mathrm{ext}}^{21} = \left|\frac{\varphi_{\mathrm{ext}}^{21} - \varphi_1}{\varphi_{\mathrm{ext}}^{12}}\right| \qquad \text{[B.4]}$$

Finally, the fine-grid convergence index is given by:

$$GCI_{\mathrm{fine}}^{21} = \frac{1.25 e_{a}^{21}}{r_{21}^{p} - 1} \qquad \text{[B.5]}$$

**Table B.1** presents two examples of calculations for the pressure drop on the jumper. The case considered is 3D simulations with the Realizable $\kappa$-$\varepsilon$ model, applied for $U^0_{GS}$ = 10 m/s, $\varepsilon_{i\text{-}hor}$ = 50%, $t_{ramp}$ = 2s. As shown in the table, the numerical uncertainty in the fine-grid solution for the pressure drop at $t$=2.35 s and 4.2 s are 5.11% and 4.49%. **Figure B.1** shows the pressure drop on the jumper for this case. The error bars at ($b$) indicate the numerical uncertainty calculated by **Eq. [B.5]**. The local order of accuracy, $p$, calculated from **Eq. [B.2]** ranges from 0.479 to 8.88, with a global average value of 3.07. Oscillatory convergence occurs at 16% of the points. The maximal discretization uncertainty is 14%, which corresponds to $t$ = 2.85 s.

**Table B.1:** Sample calculation of discretization error for the pressure drop on the jumper. The simulated case is 3D model: Realizable $\kappa$-$\varepsilon$, $U^0_{GS}$ = 10 m/s, $\varepsilon_{i\text{-}hor}$ = 50%, $t_{ramp}$ = 2s.

| **$\varphi$ is the pressure drop on the jumper** | | |
|---|---|---|
| $N_1$; $N_2$; $N_3$ | 5,956,875; 3,071,250; 1,292,400 | |
| $r_{21}$ | 1.39 | |
| $r_{32}$ | 1.54 | |
| $t$ [s] | 2.35 | 4.2 |
| $\varphi_1$ [Pa] | 4634.63 | 1784.13 |
| $\varphi_2$ [Pa] | 4429.93 | 1660.72 |
| $\varphi_3$ [Pa] | 3797.66 | 1086.2 |
| $\varepsilon_{32}$ | -632.26 | -574.513 |
| $\varepsilon_{21}$ | -204.69 | -123.41 |
| $\varepsilon_{32}/\varepsilon_{21}$ | 3.08 | 4.65 |
| $p$ | 2.21 | 3.23 |
| $\varphi_{ext}^{21}$ [Pa] | 4824.14 | 1848.28 |
| $e_a^{21}$ [%] | 4.41 | 6.91 |
| $e_{ext}^{21}$ [%] | 3.92 | 3.47 |
| $GCI_{fine}^{21}$ [%] | 5.11 | 4.49 |

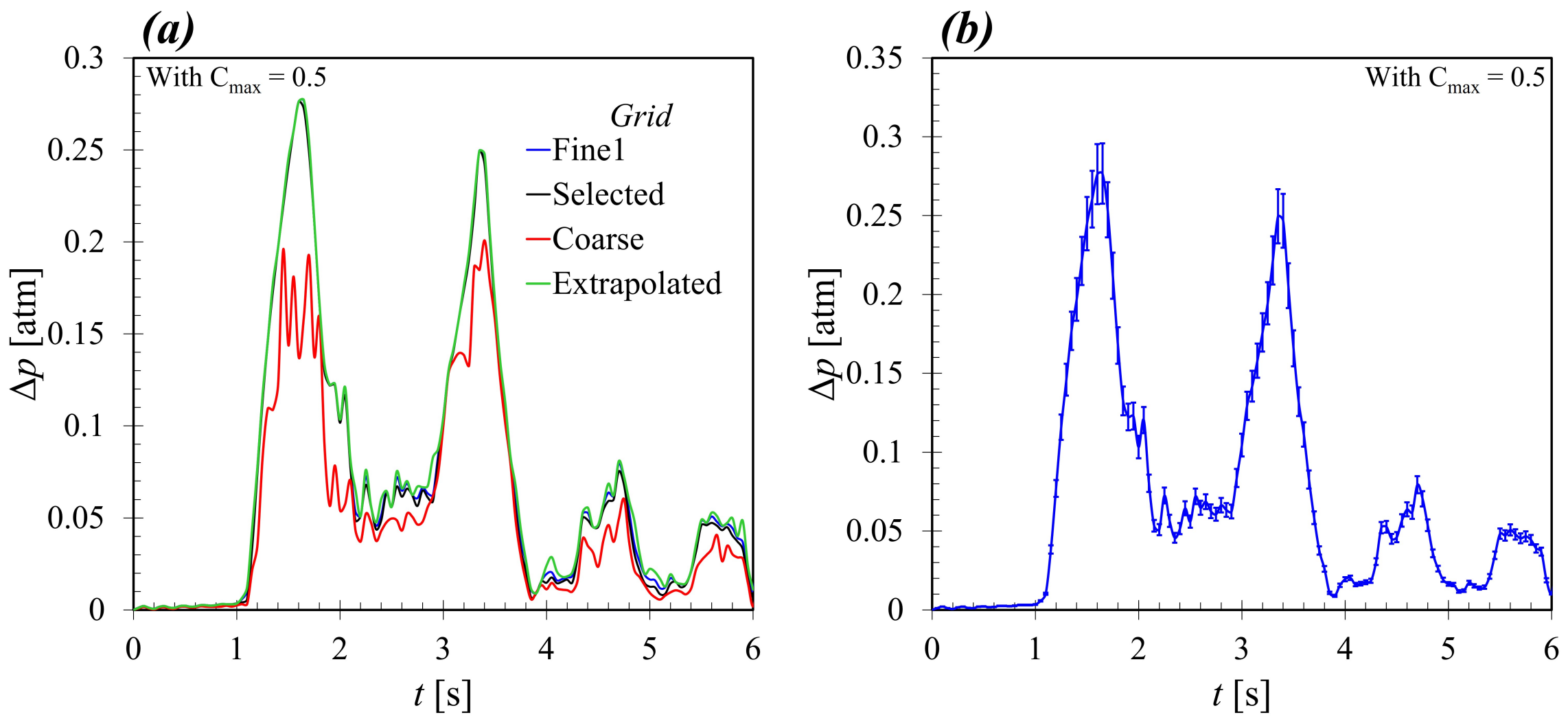

**Figure B.1:** Pressure drop obtained in 3D simulations with the Realizable $\kappa$-$\varepsilon$ model, $U^0_{GS}$ = 10 m/s, $\varepsilon_{i\text{-}hor}$ = 50%, $t_{ramp}$ = 2 s case. ($a$) Extrapolated values and solutions obtained by different grids; ($b$) Fine grid solution with discretization error bars computed by **Eq. [B.5]**

**Table B.2** presents the discretization error estimation for the horizontal force obtained for the same case. Note that the force is ~zero for most of the simulated time and it exhibits an instant peak (which is the interest of the present study), and the CGI calculation procedure does not work for points with $\varepsilon_{32}$ or $\varepsilon_{21}$ that are close to zero. The examined value corresponds to the maximal force peak applied on the riser's upper elbow. As shown in the table, the numerical uncertainty in the fine-grid solution for the horizontal force at $t$=1.9 s is 5.81%.

**Table B.2:** Sample calculation of discretization error for the maximal horizontal force applied on the riser's upper elbow. The simulated case is 3D model: Realizable $\kappa$-$\varepsilon$, $U^0_{GS}$ = 10 m/s, $\varepsilon_{i\text{-}hor}$ = 50%, $t_{ramp}$ = 2s.

| **$\varphi$ is the horizontal force applied on the riser's upper elbow** | |
|---|---|
| $N_1$; $N_2$; $N_3$ | 5,956,875; 3,071,250; 1,292,400 |
| $r_{21}$ | 1.39 |
| $r_{32}$ | 1.54 |
| $t$ [s] | 1.9 |
| $\varphi_1$ [N] | 174.29 |
| $\varphi_2$ [N] | 165.15 |
| $\varphi_3$ [N] | 136.15 |
| $\varepsilon_{32}$ | -29 |
| $\varepsilon_{21}$ | -9.13 |
| $\varepsilon_{32}/\varepsilon_{21}$ | 3.173 |
| $p$ | 2.27 |
| $\varphi_{ext}^{21}$ [N] | 182.39 |
| $e_a^{21}$ [%] | 5.24 |
| $e_{ext}^{21}$ [%] | 4.44 |
| $GCI_{fine}^{21}$ [%] | 5.81 |